\documentclass{birkjour}
\usepackage{latexsym}
\usepackage{amsfonts} 
\usepackage{ascmac}
\usepackage[mathscr]{euscript}
\usepackage{amsmath,mathrsfs,bm,amssymb,color}
\newtheorem{define}{Definition}[section]
\newtheorem{Thm}[define]{Theorem}
\newtheorem{Prop}[define]{Proposition}
\newtheorem{lemm}[define]{Lemma}
\newtheorem{rem}[define]{Remark}

\newtheorem{example}{Example}
\newtheorem{coro}[define]{Corollary}

\newcommand{\one}{{\mathchoice {\rm 1\mskip-4mu l} {\rm 1\mskip-4mu l}
{\rm 1\mskip-4.5mu l} {\rm 1\mskip-5mu l}}}
\newcommand{\h}{\mathfrak{h}}

\newcommand{\ex}{\mathrm{e}}
\newcommand{\D}{\mathrm{dom}}

\newcommand{\im}{\mathrm{i}}
\newcommand{\Fock}{\mathfrak{F}}

\newcommand{\dG}{d\Gamma}

\newcommand{\ran}{\mathrm{ran}}
\newcommand{\AFock}{\mathfrak{F}_{\mathrm{e}}}
\newcommand{\la}{\langle}
\newcommand{\ra}{\rangle}
\newcommand{\Tr}{\mathrm{Tr}}

\newcommand{\BbbR}{\mathbb{R}}
\newcommand{\BbbN}{\mathbb{N}}
\newcommand{\BbbZ}{\mathbb{Z}}

\newcommand{\BbbC}{\mathbb{C}}
\newcommand{\vepsilon}{\varepsilon}
\newcommand{\vphi}{\varphi}

\newcommand{\no}{\nonumber \\}
\renewcommand{\labelenumi}{\rm{(\roman{enumi})}}

\newcommand{\bphi}{\boldsymbol q}
\newcommand{\bomega}{\boldsymbol \omega}
\newcommand{\Ue}{U_{\mathrm{eff}}}
\newcommand{\ef}{\mathrm{eff}}
\newcommand{\Md}{M^{\dagger}}
\newcommand{\mb}{\mathbf}

\newcommand{\bfphi}{{\boldsymbol \vphi}}

\newcommand{\tU}{\widetilde{\mathbf{U}}}
\newcommand{\Sig}{{\boldsymbol \sigma}}
\newcommand{\balpha}{{\boldsymbol \alpha}}

\begin{document}

%-------------------------------------------------------------------------
% editorial commands: to be inserted by the editorial office
%
%\firstpage{1} \volume{228} \Copyrightyear{2004} \DOI{003-0001}
%\seriesextraline{This is the Concrete Title of this Book\br H.E. R and S.T.C. W, Eds.}
%\firstpage{1}

\title[
 Rigorous results concerning the Holstein--Hubbard model
]
 {
 Rigorous results concerning 
the Holstein--Hubbard model
}

%----------Author 1
\author[Tadahiro Miyao]{Tadahiro Miyao}

\address{%
 Department of Mathematics,\\
 Hokkaido University,\\
 Sapporo 060-0810, Japan
}

\email{
miyao@math.sci.hokudai.ac.jp
}

%\subjclass{Primary 99Z99; Secondary 00A00}

%\keywords{Class file, journal}

\date{January 1, 2004}

\begin{abstract}
The Holstein model has been widely accepted as a model comprising   electrons
 interacting with  phonons;  analysis of this model's  ground states was
 accomplished two 
 decades ago. However,  the  results were obtained without  completely
 taking  
  repulsive Coulomb interactions into account. 
Recent progress has made it possible to treat such interactions
 rigorously; 
in this paper, we study the Holstein--Hubbard model  with  repulsive
 Coulomb interactions. The ground state properties  of the model are
 investigated;  in particular, the ground state of the  Hamiltonian 
is proven to be unique for an  even number of electrons on  a bipartite
 connected lattice.
In addition, we provide  a rigorous upper bound on  charge susceptibility.
\end{abstract} 

%%% ----------------------------------------------------------------------
\maketitle
%%% ----------------------------------------------------------------------
\tableofcontents
\section{Introduction and results}\label{Intro}

\subsection{Background }

The subtle  interplay of electrons and phonons induces various physical  phenomena.
 For instance, 
when electrons interact with phonons, they have a tendency to pair.
As a result, the ground state of such a  system exhibits either superconducting
or charge-density-wave order. Another example is    high-temperature 
superconductivity. Since the discovery of coupled electron-phonon systems,  
such  systems  have become  increasingly  
active. However, a unanimously accepted  mechanism  for  the origin of high-temperature  superconductivity 
has not been established. The above-mentioned examples  suggest that 
 coupled electron-phonon systems  offer
  a rich field of study toward the identification of such a mechanism.
 In this paper, we rigorously  investigate  the  ground state properties of
 the Holstein--Hubbard model,  which is a standard model of 
 electron-phonon interaction.

The importance of  the uniqueness of ground states for models of single
particle interacting with   a  Bose field
was recognized through  rigorous studies of the quantum field theory
\cite{Faris, JFroehlich1, GlimmJaffe, Gross, Moller,  Sloan1}.
Field-theoretical methods have been successfully adopted in 
condensed matter physics.
In particular, L\"owen \cite{Lowen} applied Fr\"ohlich's method
\cite{JFroehlich1} to a model of a single electron  positioned 
on a discrete lattice system and  that interacts with the phonons of lattice.
Recently, this method was  extended to  a two-electron system
interacting with  phonons \cite{Miyao1}\footnote{
The Coulomb repulsion is considered, while the Pauli exclusion principle
is not taken into account  in \cite{Miyao1}}.

The  importance of the uniqueness of 
ground states has also been appreciated in the field of many-electron systems
 \cite{LiebMattis, LiebWu, Tasaki}. 
In addition, some  relationships  between  the  notion of correlations
and the uniqueness of the  ground states have been revealed 
 in recent years \cite{Miyao6}.
To explain why the uniqueness is important,  we recall the Hubbard model \cite{Hubbard}  as a background:
\begin{align}
H_{\mathrm{Hubbard}}=-\sum_{\{x,y\}\in E\atop{ 
\sigma\in \{\uparrow, \downarrow\}}}t_{xy}
 c_{x\sigma}^*c_{y\sigma}+\frac{1}{2}\sum_{x\in \Lambda} U_{x} (n_x-\one)^2,\ \ U_x>0.
\end{align}
For definitions of symbols, see Section \ref{DefModel}.
In general, 
the Pauli exclusion principle and the Coulomb repulsion are essential 
factors in the study of  many-electron systems.
This  model takes the  two factors  into consideration, 
and has been  regarded as a basic model of the theory of 
ferromagnetism. In \cite{Lieb}, Lieb proved the uniqueness of 
the ground state of the Hubbard model using  the method of  
spin-reflection positivity\footnote{The spin-reflection positivity
originated  from  quantum field theory \cite{OS}, 
and has various applications  to  strongly correlated electron
systems \cite{FILS, Shen, Tian}. }. 
 Ferromagnetism in the ground state immediately follows from this result.

Let us now discuss the  problem of electron-phonon interaction.
As mentioned above, the field-theoretical approach
successfully   proves the uniqueness of the  ground states of single and 
two-electron   systems interacting with phonons, however, 
 it is difficult to apply this approach to  the general many-electron systems
involving interactions  with phonons.
  Freericks and Lieb invented a crucial approach to show the uniqueness
  of  
the  many-body 
ground state of an electron-phonon Hamiltonian \cite{FL}. Their method
also
relied  on   spin-reflection positivity.   
The Freericks--Lieb  method is  applicable to a general class of  models.
To clarify the point of the argument, let us consider the Holstein
model \cite{Holstein} since it is a representative model of the  Lieb--Freericks
 class. The Hamiltonian of the Holstein model 
 is given by the following:
\begin{align}
H_{\mathrm{Holstein}}=&-\sum_{\{x,y\}\in E\atop{ 
\sigma\in \{\uparrow, \downarrow\}}}t_{xy}
 c_{x\sigma}^*c_{y\sigma}
+\sum_{x\in \Lambda}
g_x n_{x}(b_x^*+b_x)+\sum_{x\in \Lambda} \omega_0b_x^*b_x. \label{Holstein}
\end{align}
The uniqueness of the ground states of $H_{{\mathrm{Holstein}}}$
 was  successfully proved in \cite{FL}. 
As a corollary, it was  shown that the ground state has a  total spin $S=0$. 

The Holstein model  considers  the Pauli exclusion principle, but
not  Coulomb repulsion. 
It is logical as well as  important to ask whether we can prove (or disprove) the uniqueness
of the ground state even if  Coulomb repulsion is considered.
The  motivation of  this study is  to answer this question.
To investigate this problem, we analyzed the Holstein--Hubbard model
 that  contains effects of the Coulomb repulsion:
\begin{align}
H_{\mathrm{HH}}=H_{\mathrm{Holstein}}+\frac{1}{2}\sum_{x\in \Lambda} U_x
 (n_x-\one)^2,\ \ U_x>0. \label{HHPro}
\end{align} 
It should be noted  that the Lieb--Freericks approach is inapplicable to this
model\footnote{
To be precise,  their results remain  true if $U_x\le 0$,
but their method does not work if $U_x>0$.
}.
Our  first achievement is that 
we prove the uniqueness of  the ground states of the  extended Holstein--Hubbard model
 defined  by (\ref{ExtendedHH}).
As a corollary, we elucidate  the magnetic  properties of the ground state. 
To this end, we apply the theory of operator inequalities 
associated with  Hilbert cones, which has been shown to be effective in 
studies of many-electron systems \cite{Miyao1, Miyao4, Miyao5}.

At  first glance, it appears that  the form of  the Hamiltonian  is
unsuitable for application to operator inequalities because of the  electron-phonon interaction term (the middle term in the RHS of (\ref{Holstein})).
To overcome this obstacle, we employ the Lang--Firsov
transformation \cite{LF}.
By this transformation,  the electron-phonon interaction term in (\ref{HHPro})
disappears so that  we can apply our theory of operator inequalities to the resulting Hamiltonian.
This is the main reason why we use the Lang--Firosov transformation.
 Due to this transformation, the hopping matrix elements of the resulting Hamiltonian become {\it complex}-valued functions
of  the phonon coordinates [see (\ref{DefTg})].
To the best of our knowledge, there has been no attempt,  except  Miyao
\cite{Miyao2},   to show the
uniqueness of the ground states of such a Hamiltonian.
In the study by Miyao \cite{Miyao2},   the  ground state properties of the Su--Schrieffer--Heeger
 (SSH) model \cite{SSH} were
investigated.
The SSH model describes a one-dimensional  many-electron
system interacting with  phonons\footnote{
The SSH Hamiltonian is concretely given by 
\begin{align}
H_{\mathrm{SSH}}=&-\sum_{j=1}^{L}\sum_{
\sigma\in \{\uparrow, \downarrow\}}(q_{j}-q_{j+1})t
 c_{j\sigma}^*c_{j+1\sigma}
+
\frac{1}{2}\sum_{j=1}^L U_j (n_j-\one)^2
+\sum_{j =1}^L \omega_0b_j^*b_j,
\end{align}
where $q_j=b_j-b_j^*$ and $t> 0$. Clearly,  the hopping matrix element 
$t_{j}({\bf q}):=-(q_j-q_{j+1}) t$ depends on  phonon coordinates, 
$\{q_j\}_{j\in \Lambda}$.
}. 
A significant  feature of this model is that  its hopping matrix
elements are {\it real}-valued functions of the phonon coordinates which
makes our analysis complicated.
Since the elements of the hopping matrix are complex in our case,
 the method in \cite{Miyao2}  cannot be  applied directly. 
 Therefore, we 
establish a  more   sophisticated 
analysis in this study.

 Lieb's results for the Hubbard model concern the ground state.
On the other hand, Kubo and Kishi showed a finite temperature version of
Lieb's theorem \cite{KuboKishi}.
They showed a uniform upper bound on the charg susceptibility of the Hubbard model 
at finite temperature, which implies  the absence of  charge long-range order.
As the second achievement of this study, we extend their result to the
extended Holstein--Hubbard model.

Our method requires a  restriction on the electron-phonon coupling
strength ($|g_0|\le \sqrt{2U_0/\omega_0}$).
We are aware of 
no rigorous results when the electron-phonon coupling strength is large enough ($|g_0|> \sqrt{2U_0/\omega_0}$).

\subsection{The extended Holstein--Hubbard model}\label{DefModel}
Let $G=(\Lambda, E)$ be a graph with vertex set $\Lambda$ and edge
collection $E$. We suppose that $G$ is embedded in $\BbbR^d$ and that 
$\Lambda$ is a finite subset of $\BbbR^d$. An edge
with end-points $x$ and $y$ is  denoted by $\{x, y\}$.
We always assume that $\{x, x\}\notin E$ for any $x\in
\Lambda$, i.e., any loops are excluded.
Henceforth,  we  assume that
\begin{flushleft}
 {\bf (G)} $G$ is bipartite\footnote{ A graph $G$ is called   {\it
 bipartite} if $\Lambda$ admits a partition into two classes,  such
that every edge has its ends in different classes.}.

\end{flushleft} 

The Hamiltonian of the extended  Holstein--Hubbard model is given by 
\begin{align}
H=&-\sum_{\{x,y\}\in E\atop{ 
\sigma\in \{\uparrow, \downarrow\}}}t_{xy}
 c_{x\sigma}^*c_{y\sigma}+\frac{1}{2}\sum_{x, y\in
 \Lambda}U_{xy}(n_{x}-\one)(n_y-\one)\no
&+\sum_{x, y\in \Lambda}
g_{xy}n_{x}(b_y^*+b_y)+\sum_{x\in \Lambda} \omega_0b_x^*b_x, \label{ExtendedHH}
\end{align} 
where $c_{x\sigma}$  is the electron annihilation operator at vertex $x$ and $b_x$ is
the phonon annihilation operator at vertex $x$. These operators satisfy 
the following relations:
\begin{align}
\{c_{x\sigma}, c_{x'\sigma'}^*\}=\delta_{\sigma\sigma'}\delta_{xx'},\ \ \
[b_x, b_{x'}^*]=\delta_{xx'}.
\end{align}
 $n_{x}$ is the fermionic number
 operator at vertex $x\in \Lambda$ defined by 
\begin{align}
n_x=\sum_{\sigma\in \{\uparrow, \downarrow\}} n_{x\sigma},\ \ \
 n_{x\sigma}=c_{x\sigma}^* c_{x\sigma}.
\end{align}  
$t_{xy}$ is the hopping matrix element, $U_{xy}$ is the energy of the
Coulomb interaction, and $g_{xy}$ is the strength of the electron-phonon
interaction.  We assume that 
\begin{flushleft}
{\bf (I)} $\{g_{xy}\}, \{t_{xy}\}$ and $\{U_{xy}\}$ are real symmetric
 $|\Lambda|\times |\Lambda|$ matrices\footnote{
Let $M=\{M_{xy}\}$  be a $|\Lambda|\times|\Lambda|$ matrix. $M$ is
 called a {\it real symmetric matrix}  if $M_{xy}$ is real and $M_{xy}=M_{yx}$ for
 all $x,y\in \Lambda$.
}.
\end{flushleft} 
The phonons are assumed to be dispersionless with energy $\omega_0>0$.
$H$ acts in the Hilbert space 
\begin{align}
\mathfrak{E}\otimes \mathfrak{P}.
\end{align} 
$\mathfrak{E}$ is defined by $\Fock_{\mathrm{e}}\otimes
\Fock_{\mathrm{e}}$.
$\Fock_{\mathrm{e}}$ is the fermionic Fock space over $\ell^2(\Lambda)$
given by $\Fock_{\mathrm{e}}=\oplus_{n=0}^{\infty} \wedge^n
\ell^2(\Lambda)$,
where $\wedge^n \ell^2(\Lambda)$ is the $n$-fold anti-symmetric tensor
product of $\ell^2(\Lambda)$. $\mathfrak{P}$ is the bosonic Fock space
over $\ell^2(\Lambda)$ defined by $\mathfrak{P}=\oplus_{n=0}^{\infty}
\otimes_{\mathrm{s}}^n \ell^2(\Lambda)$, where $\otimes_{\mathrm{s}}^n
\ell^2(\Lambda)$ is the $n$-fold symmetric tensor product. By the
Kato--Rellich theorem,
 $H$ is
self-adjoint on $\D(N_{\mathrm{p}})$ and bounded from below\footnote{
To show  self-adjointness, recall the well-known bounds:
$
\|b_x(N_{\mathrm{p}}+\one)^{-1/2}\|\le 1,\ \ 
\|b_x^*(N_{\mathrm{p}}+\one)^{-1/2}\|\le 1
$.
Thus, we see that 
\begin{align*}\|
\sum_x g_{xy} n_x(b_y+b_y^*)\vphi
\| \le 4 |\Lambda| \max_{x,y} |g_{xy}| \|(N_{\mathrm{p}}+\one)^{1/2}\vphi\|,\
\vphi\in \D(N_{\mathrm{p}}).
\end{align*} 
Since $\|(N_{\mathrm{p}}+\one)^{1/2}\vphi\|^2\le \vepsilon
\|(N_{\mathrm{p}}+\one) \vphi\|^2+\frac{1}{4\vepsilon} \|\vphi\|^2$ for
all $\vepsilon>0$, the electron-phonon interaction term is
infinitesimally $N_{\mathrm{p}}$-bounded. Hence, we can apply the
Kato--Rellich theorem \cite{ReSi4}.

}, where
$N_{\mathrm{p}}=\sum_{x\in \Lambda} b_x^*b_x$.

Let $N_{\mathrm{e}}=\sum_{\sigma\in \{\uparrow, \downarrow\}} \sum_{x\in
\Lambda} n_{x\sigma}
$, the fermionic number operator. 
We are interested in the ground state properties of $H$ at
half-filling. Thus,  we  consider only  the following subspace: 
\begin{align}
\mathfrak{H}=\mathfrak{E}_{|\Lambda|}\otimes
 \mathfrak{P},\ \ \ 
\mathfrak{E}_{|\Lambda|}=\ker(N_{\mathrm{e}}-|\Lambda|).
\end{align} 

Let $S^{(z)}=\frac{1}{2}(N_{\mathrm{e} \uparrow}-N_{\mathrm{e}\downarrow})$, where 
$N_{\mathrm{e} \sigma}=\sum_{x\in \Lambda} n_{x\sigma},\
\sigma\in \{\uparrow, \downarrow\}$.
Since $S^{(z)}$ commutes with $H$, we have the following decompositions:
\begin{align}
\mathfrak{H}&=\bigoplus_{M=-|\Lambda|/2}^{|\Lambda|/2}\mathfrak{H}_M,\ \
 \ 
\mathfrak{H}_M=\Big(
\ker[S^{(z)}-M]\cap \mathfrak{E}_{|\Lambda|}
\Big)\otimes \mathfrak{P},\\
H&=\bigoplus_{M=-|\Lambda|/2}^{|\Lambda|/2} H_M,\ \ \ \
 H_M=H\restriction \mathfrak{H}_M.
\end{align}
Here, $\mathfrak{H}_M$ is called the $M$-subspace.
\subsection{Ground state properties} 
Before we state our first  result, we need to introduce some
definitions.

The  effective Coulomb interaction is given by the following equation:
\begin{align}
U_{\ef, xy}=U_{xy}-\frac{2}{\omega_0}\sum_{z\in \Lambda}g_{xz}g_{yz}.
\end{align}

In what follows, we assume that
\begin{flushleft}
{\bf (A. 1)}  $\displaystyle 
\sum_{x\in \Lambda} g_{xy}$ is  a constant independent of $y\in \Lambda$ .
\end{flushleft}

\begin{example}{\rm 
(i) An  example satisfying {\bf (A. 1)} is $g_{xy}=g_0\delta_{xy}$, where
 $\delta_{xy}$ is the Kronecker delta.

(ii) Let us consider a linear chain of $2L$ atoms with periodic boundary
 conditions.  In this case,  $G=(\Lambda, E)$ is defined by 
$\Lambda=\{x_j\}_{j=1}^{2L}, \ x_j\in \BbbR^2$ and $E=\{\{x_j,
 x_{j+1}\}, \{x_{j+1}, x_j\}\}_{j=1}^{2L}$ with $x_{2L+1}=x_1$.
We denote  the distance from atom $i$ to atom $j$ by $w_{i,j}=|x_i-x_j|$.
Assume that $w_{j, j+1}=\mathrm{constant}$  for all $j$. If $g_{xy}$ is a function
 of $|x-y|$, i.e., $g_{xy}=f(|x-y|)$, then {\bf (A. 1)} is satisfied. Similarly,
 if $\Lambda$ has a symmetric structure, like $\mathrm{C}_{60}$
 fullerene, then {\bf (A. 1)} is fulfilled. $\diamondsuit$
}
\end{example}

Since $G$ is bipartite, $\Lambda$ can be divided into two disjoint
sets $\Lambda_e$ and $\Lambda_o$. 
Set 
\begin{align}
\tilde{S}_+=\sum_{x\in \Lambda}  \gamma_xc_{x\uparrow} c_{x\downarrow},\ \ \
 \tilde{S}_-=\sum_{x\in \Lambda}\gamma_x c_{x\downarrow}^* c_{x\uparrow}^*,\ \ 
\tilde{S}^{(z)}=\frac{1}{2}|\Lambda|-\frac{1}{2}(N_{\mathrm{e}\uparrow}+N_{\mathrm{e}\downarrow}),
\end{align} 
where $\gamma_x=1$ for $x\in \Lambda_e$, $\gamma_x=-1$ for $x\in \Lambda_o$.
The pseudospin operator is defined by 
\begin{align}
\tilde{S}_{\mathrm{tot}}^2=\tilde{S}^{(z)2}+\frac{1}{2}\tilde{S}_+\tilde{S}_-+\frac{1}{2}\tilde{S}_-\tilde{S}_+.
\end{align} 
Although $\tilde{S}_{\mathrm{tot}}^2$ does not commute with $H_M$, it is
still useful to study ground states of $H_M$.

\begin{Thm}\label{PositiveGS}
Assume that $|\Lambda|$ is even. Assume {\bf (A. 1)}.  Assume that $\Ue$
 is positive semi-definite\footnote{
 $\Ue$ is  called {\it positive semi-definite}, if,  for
 all $\{\xi_x\}_{x\in \Lambda}\in \BbbC^{|\Lambda|}$, 
\begin{align}
\sum_{x, y\in \Lambda}\overline{\xi}_x \xi_y U_{\ef, xy}\ge 0
\end{align} 
holds. 
}. 
Then for all $M\in \{-|\Lambda|/2, -|\Lambda|/2+1, \dots, |\Lambda
|/2 \}$, among all the ground states of $H_M$, there exists at least one
 ground state $\vphi_M$ which satisfies the following:

\begin{itemize}
\item[{\rm (i)}]$\tilde{P}\vphi_M\neq 0$ holds, where $\tilde{P}$
is the orthogonal projection onto $\ker(\tilde{S}^2_{\mathrm{tot}})$.
\item[{\rm (ii)}]
 Let $S_{x+}=c_{x\uparrow}^*c_{x\downarrow}$ and
$S_{x-}=(S_{x+})^*$.  Then
\begin{align}
\big\la \vphi_M, S_{x+}S_{y-}\vphi_M\big\ra
\begin{cases}
\ge 0\ \ \mbox{if $x, y\in \Lambda_{e}$ or $x, y\in \Lambda_o$}\\
\le 0\ \ \mbox{otherwise}.\label{GSAnti}
\end{cases} 
\end{align} 
In other words, the magnetic structure of the ground state is
	     antiferromagnetic.
\end{itemize} 
\end{Thm} 
\begin{rem}{\rm
In \cite{Miyao6}, it is  pointed out  that (\ref{GSAnti}) can be regarded as the
 first Griffiths inequality.
$\diamondsuit$
}
\end{rem}
\begin{example}
{\rm 
Let  $U_{xy}=U_0\delta_{xy}$ and $g_{xy}=g_0\delta_{xy}$. Then
 $U_{\ef, xy}=(U_0-2g_0^2/\omega_0) \delta_{xy}$.
Thus,  $\Ue$ is positive semi-definite if and only if $|g_0|\le
 \sqrt{2U_0/\omega_0}$. $\diamondsuit$ 
}
\end{example} 

Theorem \ref{PositiveGS} does not exclude the  possibility
 that $H_M$ has degenerate ground states. 
Our next result concerns the uniqueness of the ground state. 
To show it, we need an additional assumption:

\begin{flushleft}
{\bf (A. 2)}  
$G$ is connected\footnote{The graph $G$ is called
 {\it connected} if any of its vertices are linked by a path in $G$.} and $t_{xy}\neq 0$ for all $\{x, y\}\in E$.
\end{flushleft}

Let us introduce the total spin operator 
\begin{align}
S_{\mathrm{tot}}^2=S^{(z)2}+\frac{1}{2} S_+S_-+\frac{1}{2}S_- S_+,
\end{align} 
where 
\begin{align}
S_+=\sum_{x\in \Lambda} c_{x\uparrow}^* c_{x\downarrow},\ \ \
 S_-=\sum_{x\in \Lambda} c_{x\downarrow}^*c_{x\uparrow}.
\end{align} 

\begin{Thm}\label{Uniqueness}
Assume that $|\Lambda|$ is even. Assume {\bf (A. 1)} and {\bf (A. 2)}.
Assume that  $\Ue$ is positive definite\footnote{
$\Ue$ will be called {\it  positive definite} if, for
 all $\{\xi_x\}_{x\in \Lambda}\in \BbbC^{|\Lambda|}\backslash\{{\bf 0}\}$, 
\begin{align}
\sum_{x, y\in \Lambda}\overline{\xi}_x \xi_y U_{\ef, xy}> 0
\end{align} 
holds. 
}. For each $M\in \{-|\Lambda|/2,
 -|\Lambda|/2+1,\dots, |\Lambda|/2\}$,
the ground state of $H_M$ is unique.  Let $\vphi_M$ be the unique ground
 state of $H_M$. Then we have the following:
\begin{itemize}
\item[{\rm (i)}] $\tilde{P}\vphi_M\neq 0$.
\item[{\rm (ii)}] There exists a unique number  $S$ such that
$S\ge |M|$ and 
	     $S_{\mathrm{tot}}^2 \vphi_M=S(S+1)\vphi_M$.
\item[{\rm (iii)}] 
\begin{align}
\big\la \vphi_M, S_{x+}S_{y-}\vphi_M\big\ra
\begin{cases}
>0\ \ \mbox{if $x, y\in \Lambda_{e}$ or $x, y\in \Lambda_o$}\\
<0\ \ \mbox{otherwise}. \label{GSAnti2}
\end{cases} 
\end{align} 
\end{itemize} 
\end{Thm} 
 \begin{rem}
{\rm 
(\ref{GSAnti2}) means that the antiferromagnetic structure becomes
  sharper than (\ref{GSAnti}) or a strict Griffiths inequality holds.
$\diamondsuit$
}
\end{rem} 
\begin{example}{\rm
Consider the case where  $U_{xy}=U_0\delta_{xy}$ and $g_{xy}=g_0\delta_{xy}$.
Then  $U_{\ef}$
 is positive definite if and only if $|g_0|<\sqrt{\omega_0
 U_0/2}$. $\diamondsuit$
}
\end{example}

\subsection{Upper bounds on  the charge  susceptibility}
We give a rigorous bound on the charge susceptibility of the
Holstein--Hubbard model. 
For simplicity, we consider the  $d$-dimensional simple cubic lattice
$\BbbZ^d$. For each $L\in \BbbN$, the vertex set is given by
\begin{align}
\Lambda=[-L, L)^d\cap \BbbZ^d.
\end{align} 
We impose a periodic boundary condition on the model. To be precise, 
the edge collection $E$ is given by
\begin{align}
E=\big\{
\{x, y\}\in \Lambda^2\, \big|\, |x-y|=1
\big\} \cup \partial,
\end{align} 
where
\begin{align}
\partial=\Big\{
\{x, y\}\in \Lambda^2\, \Big|\, |x-y|=2L-1 
\Big\}.
\end{align}
We set  $t_{xy}=t\neq 0$ for all $\{x,y\}\in E$.

Let $\delta n_x=n_x-\one$. Set
\begin{align}
\widetilde{\delta n}_p=|\Lambda|^{-1/2} \sum_{x\in \Lambda} \ex^{-\im
 x\cdot p} \delta n_x.
\end{align} 
The charge susceptibility is defined by 
\begin{align}
\chi_{\beta}(p)=\lim_{L \to \infty} \beta
 \big(\widetilde{\delta n}_{-p}, \widetilde{\delta n}_p\big)_{\beta,
 \Lambda},\ p\in [-\pi, \pi]^d,
\end{align} 
where
\begin{align}
(A, B)_{\beta, \Lambda}&=Z_{\beta, \Lambda}^{-1} \int_0^1 ds \Tr\Big[
\ex^{-s \beta( H+\sum_{x\in \Lambda}\mu_x n_x)}A \, \ex^{-(1-s)\beta (H +\sum_{x\in \Lambda}\mu_x n_x)} B
\Big],\\ 
Z_{\beta, \Lambda}&=\Tr\Big[
\ex^{-\beta(H+\sum_{x\in \Lambda}\mu_x n_x)}
\Big].
\end{align} 
The local chemical potential is given by
\begin{align}
\mu_x=\frac{2}{\omega_0} \sum_{y, z\in \Lambda} g_{xz}g_{zy}. 
\end{align} 
Note that if $g_{xy}=g_0\delta_{xy}$, then $\mu_x=2g_0^2/\omega_0$ for all
$x\in \Lambda$. 
For any $\beta$ and $\Lambda$, we can check that the thermal average
 density of the system satisfies $\la n_o\ra_{\beta, \Lambda}:=Z_{\beta,
 \Lambda}^{-1} \Tr[n_o \ex^{-\beta H}]=1$, i.e., 
 the system at half-filling is considered\footnote{
By $A:=B$, we understand that $A$ is defined in terms of $B$.
}.

We assume the following:
\begin{itemize}
\item[{\bf (B. 1)}] $g_{xy}$ and $U_{xy}$ are
	     translation-invariant, i.e., $g_{xy}=g_{x-y, o}$ and
	     $U_{xy}=U_{x-y, o}$ for all $x,y\in \Lambda$.
\item[{\bf (B. 2)}] Set $g(x)=g_{x, o}$ and $U(x)=U_{x, o}$. Then
	     $g(x)\in \ell^2(\BbbZ^d)$ and  $U(x)\in \ell^1(\BbbZ^d)$.
\item[{\bf (B. 3)}]For all $L>0$, it holds that  $\hat{U}_{\mathrm{eff}, \Lambda}(p)\ge
	     0$, where $\hat{f}_{\Lambda}(p)=\sum_{x\in \Lambda}
	     \ex^{-\im x \cdot p} f(x)$.
\end{itemize} 

\begin{rem}
{\rm 
{\bf (B. 3)} implies  that $U_{\mathrm{eff}}$ is positive semi-definite. $\diamondsuit$
}
\end{rem}

\begin{Thm}\label{HHmodel}
Assume {\bf (B. 1)}, {\bf (B. 2)}, and {\bf (B. 3)}. 
For each $p\in [-\pi, \pi]^d$
 such that $\hat{U}_{\mathrm{eff}}(p)>0$,  we have 
\begin{align}
\chi_{\beta}(p)\le \hat{U}_{\mathrm{eff}}(p)^{-1}.
\end{align}
Here $\hat{f}(p)=\sum_{x\in \BbbZ^d} \ex^{-\im x\cdot p} f(x)$.
\end{Thm} 
\begin{rem}
{\rm 
(i) By direct computation, we have
 $\hat{U}_{\mathrm{eff}}(p)=\hat{U}(p)-2\hat{g}(p)^2/\omega_0$. 

(ii) This result is an extension of the Kubo--Kishi theorem \cite{KuboKishi} in the
 following way: (a) The electron-phonon interaction is taken into
 account. (b) Not only on-site but general Coulomb repulsion is
 considered.

(iii) In a companion paper \cite{Miyao3}, we obtain a similar bound on
 the Hubbard model coupled to a quantized radiation field.
 $\diamondsuit$
}
\end{rem} 

\begin{coro}
Assume {\bf (B. 1)}, {\bf (B. 2)} and {\bf (B. 3)}. In addition, assume that  there exists a
 constant $u_0>0$ such that   $\hat{U}_{\mathrm{eff}}(p)\ge
 u_0$
for all $p\in [-\pi, \pi]^d$. Then we have  
\begin{align}
\chi_{\beta}(p)\le u_0^{-1}.
\end{align} 
Thus, by the Falk--Bruch inequality \cite{DLS, FB},  there is no charge long-range order.
\end{coro}
\begin{rem}
{\rm 
The existence of $u_0>0$ implies that $\Ue$ is positive definite. $\diamondsuit$
}
\end{rem} 

\begin{example}
{\rm  
For each $U_0, U_1, g_0\ge 0$, let 
\begin{align}
U_{xy}=\begin{cases}
 U_0 & x=y\\
U_1/2d & |x-y|=1\\
0 & \mbox{otherwise}
\end{cases}, \hspace{2cm}
g_{xy}=g_0\delta_{xy}.
\end{align} 
Clearly, {\bf (B. 1)} and {\bf (B. 2)} are satisfied.
Then one sees
 $\hat{U}_{\mathrm{eff}}(p)=(U_0-U_1-2g_0^2/\omega_0)+\frac{U_1}{d}
 \sum_{j=1}^d(1+\cos p_j)$. Thus, {\bf (B. 3)} is satisfied whenever
$U_0-U_1-2g_0^2/\omega_0\ge 0$. 
There is no charge long-range order if $U_0-U_1-2g_0^2/\omega_0> 0$.
If $U_0-U_1-2g_0^2/\omega_0=0$, then $\chi_{ \beta}(p)$
 could diverge at extreme points of $[-\pi, \pi]^d$.
$\diamondsuit$
}
\end{example} 

\begin{rem}
{\rm 
In the case where $U_0-U_1-2g^2/\omega_0<0$, the existence of charge
 long-range order is proved in \cite{Miyao7}. $\diamondsuit$
 
}
\end{rem}

\subsection{Organization}
The organization of the paper is as follows: In Section \ref{Prel}, we
introduce several operator inequalities related to  Hilbert
cones. These operator inequalities are very useful for our study.  
Sections \ref{Sec3}-\ref{InfB} are  devoted to proving  the  main results in Section \ref{Intro}.

In Section \ref{Sec3}, we provide several expressions of the Hamiltonian (\ref{ExtendedHH})
 by performing the hole-particle and Lang--Firsov transformations.
We then choose a suitable expression in each 
section below.

In Section \ref{Sec4}, we show  Theorem 
\ref{PositiveGS}. 
By choosing a suitable Hilbert cone, we prove that the heat semi-group
 generated by the Hamiltonian preserves the positivity.
Theorem \ref{PositiveGS} is a corollary of this fact.

In Section  \ref{Sec5}, proof of  Theorem \ref{Uniqueness}  is  given. 
We show that the semi-group generated by the Hamiltonian
improves the positivity with respect to the  Hilbert cone constructed in 
 Section \ref{Sec4}. The uniqueness of ground states follows from
 Faris' theorem, which is a generalization of the Perron--Frobenius
 theorem.
By applying this fact, the some magnetic structures of the ground state are
revealed.

Section  \ref{InfB} is devoted to  the proof of Theorem
\ref{HHmodel}. 
We obtain an upper bound on the charge susceptibility by extending the
method
 of Gaussian domination established in \cite{DLS,FILS,FSS}.

In  Appendices   \ref{GeneralSC} and \ref{SPI}, we give a list
of basic facts that are used  in  the main  sections.

In Appendix \ref{PfC}, we give a proof of a technical proposition which is needed in Section \ref{Sec5}.

\begin{flushleft}
{\bf Acknowledgements.}
This work was supported by KAKENHI(20554421).
I would be grateful to  the anonymous referees for useful comments.
\end{flushleft} 

\section{Preliminaries}\label{Prel}

\subsection{Hilbert cones and their associated operator inequalities}
\begin{define}\label{HilCone}
{\rm
Let $\mathfrak{X}$ be a complex Hilbert space. 
By a {\it convex  cone}, we denote a closed convex set
 $\mathfrak{X}_+ \subseteq \mathfrak{X}$
such that $t\mathfrak{X}_+ \subseteq \mathfrak{X}_+$ for all $t\ge 0$ and $\mathfrak{X}_+\cap (-\mathfrak{X}_+)=\{0\}$. In what follows, we always assume that $\mathfrak{X}_+\neq\{0\}$.
 A convex cone,
 $\mathfrak{X}_+$ in $\mathfrak{X}$, is called  a {\it Hilbert cone} if it
 satisfies
 the following\footnote{$\mathfrak{X}_+$ is a Hilbert cone if and only
 if $\mathfrak{X}_+$ is a self-dual cone \cite{Bos, Miyao1, Miyao6}.}: 
\begin{itemize}
\item[(i)] $ \la x, y\ra\ge 0$ for all $x, y\in \mathfrak{X}_+$.
\item[(ii)] Let $\mathfrak{X}_{\BbbR}$
 be a real subspace of $\mathfrak{X}$ generated by $\mathfrak{X}_+$ . Then
	     for all $x\in \mathfrak{X}_{\BbbR}$, there exist $x_+,
	     x_-\in \mathfrak{X}_+$ such that $x=x_+-x_-$ and $\la x_+,
	     x_-\ra=0$.
\item[(iii)] $\mathfrak{X}=\mathfrak{X}_{\BbbR}+\im
	    \mathfrak{X}_{\mathbb{R}}= \{x+\im y\, |\, x, y\in \mathfrak{X}_{\BbbR}\}$.

\end{itemize} 
A vector $x$ is said to be  {\it positive w.r.t. $\mathfrak{X}_+$} if $x\in
 \mathfrak{X}_+$.  We write this as $x\ge 0$  w.r.t. $\mathfrak{X}_+$.

 A vector $y\in \mathfrak{X}$ is called {\it strictly positive
w.r.t. $\mathfrak{X}_+$} whenever $\la x, y\ra>0$ for all $x\in
\mathfrak{X}_+\backslash \{0\}$. We write this as $x>0 $
w.r.t. $\mathfrak{X}_+$. $\diamondsuit$
}
\end{define} 
In  subsequent  sections, we will use the following operator inequalities:
\begin{define}{\rm 
We denote by  $\mathscr{B}(\mathfrak{X})$  the set of all bounded linear operators on
$\mathfrak{X}$.
Let $A, B\in \mathscr{B}(\mathfrak{X})$.
\begin{itemize}
\item[{\rm (i)}] If $A \mathfrak{X}_+\subseteq \mathfrak{X}_+$\footnote{
For each subset $\mathfrak{Y}\subseteq \mathfrak{X}$, $A\mathfrak{Y}$ is
	     defined by $A\mathfrak{Y}=\{Ax\, |\, x\in \mathfrak{Y}\}$.
}, we then 
write  this as  $A \unrhd 0$ w.r.t. $\mathfrak{X}_+$\footnote{This
 symbol was introduced by Miura \cite{Miura}, see also \cite{KiRo2}.}. In
	     this case, we say that {\it $A$ preserves the
positivity w.r.t. $\mathfrak{X}_+$.}  Suppose that $A\mathfrak{X}_{\BbbR}\subseteq
 \mathfrak{X}_{\BbbR}$ and $B\mathfrak{X}_{\BbbR} \subseteq
	     \mathfrak{X}_{\BbbR}$. If $(A-B) \mathfrak{X}_+\subseteq
	     \mathfrak{X}_+$, then we write this as $A \unrhd B$ w.r.t. $\mathfrak{X}_+$.
 \item[{\rm (ii)}]
We write  $A\rhd 0$ w.r.t. $\mathfrak{X}_+$, if  $Ax >0$ w.r.t. $\mathfrak{X}_+$ for all $x\in
\mathfrak{X}_+ \backslash \{0\}$. 
 In this case, we say that {\it $A$ improves the
positivity w.r.t. $\mathfrak{X}_+$.} $\diamondsuit$
\end{itemize} 
}
\end{define} 
 
The following proposition is fundamental to this paper:

\begin{Prop}
Let $A, B, C, D\in \mathscr{B}(\mathfrak{X})$ and let $a, b\in
 \BbbR$. We have  the following:
\begin{itemize}
\item[{\rm (i)}] If $A\unrhd 0, B\unrhd 0$ w.r.t. $\mathfrak{X}_+$ and
	     $a, b\ge 0$, then $aA +bB \unrhd 0$
	     w.r.t. $\mathfrak{X}_+$.
\item[{\rm (ii)}] If $A \unrhd B \unrhd 0$ and $C\unrhd D \unrhd 0$
	     w.r.t. $\mathfrak{X}_+$,
	     then
$AC\unrhd BD \unrhd 0$ w.r.t. $\mathfrak{X}_+$.
\end{itemize} 
\end{Prop} 
{\it Proof.} (i) is trivial.

(ii) If $X\unrhd 0$ and $Y\unrhd 0$
w.r.t. $\mathfrak{X}_+$, 
we have $XY\mathfrak{X}\subseteq X\mathfrak{X} \subseteq \mathfrak{X}$.
Hence,  it holds that $XY\unrhd 0$
w.r.t. $\mathfrak{X}_+$.
Hence, we have 
\begin{align*}
AC-BD=\underbrace{A}_{\unrhd 0}\underbrace{(C-D)}_{\unrhd 0}+
 \underbrace{(A-B)}_{\unrhd 0} \underbrace{D}_{\unrhd 0} \unrhd 0\ \ \
 \mbox{w.r.t. $\mathfrak{X}_+$}.
\end{align*} 
This completes the proof. $\Box$
\medskip\\

In Appendix \ref{GeneralSC}, we give several crucial theorems on the operator
inequalities associated with Hilbert cones. 

\subsection{A canonical cone in $\mathscr{L}^2(\mathfrak{h})$}

Let $\mathfrak{h}$ be a complex Hilbert space. The set of all
Hilbert-Schmidt class operators on $\mathfrak{h}$ is denoted  by
$\mathscr{L}^2(\mathfrak{h})$, i.e.,  
$
\mathscr{L}^2(\mathfrak{h})=\{
\xi\in \mathscr{B}(\mathfrak{h})\, |\, \Tr[\xi^* \xi]<\infty
\}$.
Henceforth, we regard $\mathscr{L}^2(\mathfrak{h})$ as a Hilbert space equipped
with  the inner product $\la \xi, \eta \ra_{\mathscr{L}^2}=\Tr[\xi^*
\eta],\,   \xi, \eta\in \mathscr{L}^2(\mathfrak{h})$. 
For each $A\in \mathscr{B}(\mathfrak{h})$, the left multiplication
operator is defined by
\begin{align}
\mathcal{L}(A)\xi=A\xi,\ \ \xi\in \mathscr{L}^2(\mathfrak{h}).
\end{align} 
Similarly, the right multiplication operator is defined by 
\begin{align}
\mathcal{R}(A)\xi=\xi A, \ \ \xi\in \mathscr{L}^2(\mathfrak{h}).
\end{align} 
It is not  hard to check that 
\begin{align}
\mathcal{L}(A)\mathcal{L}(B)=\mathcal{L}(AB),\ \
 \mathcal{R}(A)\mathcal{R}(B)=\mathcal{R}(BA),\ \ A, B\in \mathscr{B}(\mathfrak{h}).
\end{align} 
\begin{define}{\rm 
A canonical   cone in $\mathscr{L}^2(\mathfrak{h})$ is given by
\begin{align}
\mathscr{L}^2(\mathfrak{h})_+= \Big\{\xi\in \mathscr{L}^2(\mathfrak{h})\, \Big|\,\mbox{$\xi$ is
 self-adjoint and $\xi\ge 0$
  as
 an operator on $\mathfrak{h}$} \Big\}.
\end{align} 
(Recall that a linear operator $\xi$ on $\h$ is said to be positive if $\la x,
\xi x\ra_{\h} \ge 0$ for all $x\in \h$. We write this as $\xi\ge 0$.)
$\diamondsuit$
}
\end{define} 

\begin{Prop}\label{SDL2}
$\mathscr{L}^2(\h)_+$ is a Hilbert cone in $\mathscr{L}^2(\h)$.
\end{Prop} 
{\it Proof.}
 We will check  conditions (i)-(iii) in Definition \ref{HilCone}.

(i) Let $\xi, \eta\in \mathscr{L}^2(\h)_+$. Since $\xi^{1/2} \eta \xi^{1/2}\ge
0$, we have $\la \xi, \eta\ra_{\mathscr{L}^2}=\Tr[\xi\eta]=\Tr[\xi^{1/2}\eta
\xi^{1/2}] \ge 0$.

(ii) Note that $\mathscr{L}^2(\h)_{\BbbR}=\{\xi\in \mathscr{L}^2(\h)\, |\, \mbox{$\xi$ is
self-adjoint }\}$. Let $\xi\in \mathscr{L}^2(\h)_{\BbbR}$. By the spectral
theorem,
there  is a projection valued measure $\{E(\cdot)\}$ such that
$\xi=\int_{\BbbR} \lambda dE(\lambda)$.
Denote $
\xi_+=\int_0^{\infty} \lambda dE(\lambda)
$
and 
$
\xi_-=\int_{-\infty}^0 (-\lambda)dE(\lambda)
$. Clearly, it holds that $\xi_+\xi_-=0, \xi_{\pm}\in \mathscr{L}^2(\h)_+$ and
$\xi=\xi_+-\xi_-$. Thus, (ii) is satisfied.

(iii) For each $\xi\in \mathscr{L}^2(\h)$, we have $\xi=\xi_{R}+\im \xi_I$, where 
$\xi_R=(\xi+\xi^*)/2$ and  $\xi_I=(\xi-\xi^*)/2\im$. Trivially, $\xi_R,
\xi_I\in \mathscr{L}^2(\h)_{\BbbR}$. This completes the proof. $\Box$

\begin{lemm}\label{GeneralPP}
Let $A\in \mathscr{B}(\mathfrak{h})$. We have 
 $\mathcal{L}(A^*)\mathcal{R}(A)\unrhd 0$ w.r.t. $\mathscr{L}^2(\mathfrak{h})_+$.
\end{lemm} 
{\it Proof.} For each $\xi\in \mathscr{L}^2(\mathfrak{h})_+$,  we have 
$
\mathcal{L}(A^*)\mathcal{R}(A)\xi=A^*\xi A \ge 0.
$ $\Box$

\section{Several expressions of the Hamiltonian,  $H$}\label{Sec3}

\subsection{The Lang--Firsov transformation} \label{SecLF}

Let
\begin{align}
q_x=\frac{1}{\sqrt{2\omega_0}}(b_x^*+b_x),\ \ \ p_x=\im \sqrt{\frac{\omega_0}{2}}(b_x^*-b_x).
\end{align} 
Both operators are essentially self-adjoint. We denote their closures by
the same symbols.
Let 
\begin{align}
L=-\im  \sqrt{2}\omega_0^{-3/2} \sum_{x, y\in \Lambda}g_{xy}n_x p_y.
\end{align} 
 $L$ is essentially anti-self-adjoint.
We also denote its closure by the same symbol. Hence,   $\ex^L$ is a unitary 
operator\footnote{The unitary operator $\ex^{L}$ was introduced by Lang
and Firsov \cite{LF}.}.  We  see that 
\begin{align}
\ex^{L}c_{x\sigma}\ex^{-L}&=\exp\Bigg\{
\im \sqrt{2}\omega_0^{-3/2} \sum_{y\in \Lambda}g_{xy}p_y
\Bigg\}c_{x\sigma}, \label{LFTC}\\ 
\ex^{L}b_x\ex^{-L}
&=b_x-\omega_0^{-1}\sum_{y\in \Lambda}g_{yx}n_y. \label{LFT}
\end{align} 

Let
\begin{align}
V_{xy}=\sum_{z\in \Lambda} \frac{2}{\omega_0}g_{xz}g_{yz}.
\end{align}

Using the facts that 
 \begin{align}
\ex^{-\im \frac{\pi}{2}N_{\mathrm{p}}} q_x \ex^{\im
 \frac{\pi}{2}N_{\mathrm{p}}}
=\omega_0^{-1}p_x,\ \ \ \ex^{-\im \frac{\pi}{2}N_{\mathrm{p}}} p_x \ex^{\im
 \frac{\pi}{2}N_{\mathrm{p}}}
=\omega_0 q_x, \label{PQ}
\end{align}
where $
N_{\mathrm{p}}=\sum_{x\in \Lambda} b_x^* b_x
$,  one arrives at the following:

\begin{Prop}\label{LangFarisov}
Set $\mathscr{U}=\ex^{-\im \frac{\pi}{2}N_{\mathrm{p}}} \ex^L$. 
We define $\widehat{H}_M$ by 
\begin{align}
\widehat{H}_M=\mathscr{U}H_M \mathscr{U}^*-\frac{1}{2}\sum_{x, y\in \Lambda}
 V_{xy}+\frac{g_*^2}{\omega_0^2}(|\Lambda|-2M),
\end{align}  where $g_*=\sum_{x\in
 \Lambda} g_{xy}$\footnote{By {\bf (A. 1)}, $g_*$ is a constant
 independent of $y$. }.
Then  we have 
\begin{align}
\widehat{H}_M=-T_{-g,\uparrow}-T_{-g, \downarrow} +H_{\mathrm{p}}+\mathbf{U},
\end{align}
where 
\begin{align}
T_{\pm g, \sigma}=&\sum_{\{x, y\}\in E}
t_{xy} c_{x\sigma}^* c_{y\sigma} \exp\Big\{
\pm \im \Phi_{\{x, y\}}\Big\}\label{DefTg}
,\\
\Phi_{\{x, y\}}=& \sqrt{2}\omega_0^{-1/2}
\sum_{z\in \Lambda} 
(
g_{xz}-g_{yz}
)
q_z,\label{Phase}
\\
H_{\mathrm{p}}=&\frac{1}{2}\sum_{x\in \Lambda}\big(p_x^2+\omega_0^2 q_x^2\big),\\
\mathbf{U}=&\frac{1}{2}\sum_{x, y\in \Lambda} U_{\ef, xy}(n_{x}-\one\big)\big( n_{y\,
}-\one\big),\\
U_{\ef, xy}=& U_{xy}-V_{xy}.
\end{align}
\end{Prop} 
{\it Proof.} We note the following:
\begin{align}
\sum_{x, y\in \Lambda} V_{xy}n_{y\sigma}= \frac{2}{\omega_0}g_*^2
 N_{\mathrm{e}\sigma}=\frac{1}{\omega_0}g_*^2(|\Lambda|-2M)\ \ \
 \mbox{on $\mathfrak{H}_M$}.
\end{align} 
Here, we used {\bf (A. 1)}.
Thus,  the formula immediately follows from  (\ref{LFTC}), (\ref{LFT})
and (\ref{PQ}). $\Box$ 

\subsection{Expression  of the Hamiltonian in
  $(\Fock_{\mathrm{e}}\otimes \Fock_{\mathrm{e}})\otimes
  \mathfrak{P}$}

Note that
\begin{align}
c_{x\uparrow}=c_x\otimes \one,\ \ \ c_{x\downarrow}=(-\one)^{\mathsf{N}_{\mathrm{e}}}\otimes c_x,
\end{align} 
where $c_x$ and $c_x^*$ are the fermionic annihilation- and creation
operators on $\AFock$, and $\mathsf{N}_{\mathrm{e}}$ is the fermionic number
operator given by  $\mathsf{N}_{\mathrm{e}}=\sum_{x\in \Lambda}\mathsf{n}_x$ with $\mathsf{n}_x=c_x^*c_x$.
Thus,  we have the following:
\begin{align}
T_{\pm g, \uparrow}&=\sum_{\{x, y\}\in E} t_{xy}c_x^* c_y\otimes \one
 \otimes \exp\Big\{
\pm \im \Phi_{\{x, y\}}\Big\},\\
T_{\pm g, \downarrow}&=\sum_{\{x, y\}\in E} t_{xy}\one \otimes c_x^* c_y\otimes  \exp\Big\{
\pm \im \Phi_{\{x, y\}}\Big\},\\
\mathbf{U}&=\frac{1}{2}\sum_{x, y\in \Lambda} U_{\ef,
 xy}\big(\mathsf{n}_x\otimes \one +\one \otimes \mathsf{n}_x-\one
 \big)\big(\mathsf{n}_y\otimes \one +\one \otimes \mathsf{n}_y-\one
 \big)\otimes \one_{\mathfrak{P}},
\end{align} 
where $\one_{\mathfrak{P}}$ is the identity operator on $\mathfrak{P}$.

\subsection{The hole-particle transformation}\label{SecHP}

The {\it hole-particle transformation} is a unitary operator $\mathcal{W}$ on $\mathfrak{E}_{|\Lambda|}$
such that  
\begin{align}
&\mathcal{W}c_{x}\otimes \one \mathcal{W}^*=\gamma_x
 c_{x}^*\otimes \one,\ \ \  
\mathcal{W}c_{x}^*\otimes \one  \mathcal{W}^*=\gamma_x
 c_{x}\otimes \one,\ \ 
\mathcal{W}\one \otimes  c_{x}\mathcal{W}^* =\one \otimes c_{x }.
\end{align}

Observe that $\mathcal{W}
N_{\mathrm{e}} \mathcal{W}^*= |\Lambda|- (\mathsf{N}_{\mathrm{e}}\otimes \one -\one
\otimes \mathsf{N}_{\mathrm{e}})$ and $\mathcal{W}S^{(z)}
\mathcal{W}^*=\frac{1}{2}|\Lambda|-\tfrac{1}{2}(\mathsf{N}_{\mathrm{e}}\otimes \one +\one
\otimes \mathsf{N}_{\mathrm{e}})$. Hence, 
we  have 
\begin{align}
\mathcal{W} \mathfrak{E}_{|\Lambda|}=\bigoplus_{n=0}^{|\Lambda|}
 \Fock_{\mathrm{e}, n}\otimes \Fock_{\mathrm{e}, n},\ \ \   \mathcal{W}
\mathfrak{H}_M=\Fock_{\mathrm{e}, (|\Lambda|-2M)/2} \otimes
 \Fock_{\mathrm{e}, (|\Lambda|-2M)/2},
\end{align} 
where $\Fock_{\mathrm{e}, n}=\wedge^n \ell^2(\Lambda)$.
In what follows, we set 
\begin{align}
\Md=\frac{1}{2}(|\Lambda|-2M).
\end{align}

\begin{lemm}
We have the following:
\begin{itemize}
\item[{\rm (i)}] $\displaystyle \mathcal{W} T_{-g, \uparrow} \mathcal{W}^* =T_{+g,
	     \uparrow}.$
 \item[{\rm (ii)}]$\mathcal{W} T_{-g, \downarrow}
	     \mathcal{W}^* =T_{-g, \downarrow}$.
\item[{\rm (iii)}] $\mathcal{W} \mathbf{U}
	     \mathcal{W}^*=\widetilde{\mathbf{U}}$, where
\begin{align}
\widetilde{\mathbf{U}}= \frac{1}{2}\sum_{x, y\in \Lambda} U_{\ef, xy}
\big(\mathsf{n}_x\otimes \one -\one \otimes \mathsf{n}_x
 \big)\big(\mathsf{n}_y\otimes \one -\one \otimes \mathsf{n}_y
 \big)\otimes \one_{\mathfrak{P}}.
\end{align} 
\end{itemize} 
\end{lemm} 
{\it Proof.}
(i) 
By definition of $\mathcal{W}$, we have 
\begin{align}
&\mathcal{W}
\sum_{\{x, y\}\in E} t_{xy}
 c_x^* c_y \otimes \one \otimes \exp\Big\{
-\im \Phi_{\{x, y\}}
\Big\}
\mathcal{W}^*\no
=&\sum_{\{x, y\}\in E} t_{xy}
 \gamma_x \gamma_y c_x c_y^* \otimes \one \otimes \exp\Big\{
-\im \Phi_{\{x, y\}}
\Big\}.\label{WTW}
\end{align} 
Since $G$ is bipartite, $\gamma_x\gamma_y=-1$ holds for all $\{x,
y\}\in E$. Consequently, 
\begin{align}
\mbox{RHS of (\ref{WTW})}&=\sum_{\{x, y\}\in E} t_{xy}
 c_y^* c_x \otimes \one \otimes \exp\Big\{
-\im \Phi_{\{x, y\}}
\Big\}\\
&=\sum_{\{y, x\}\in E} t_{yx}
 c_x^* c_y \otimes \one \otimes \exp\Big\{
-\im \Phi_{\{y, x\}}
\Big\}\no
&=\sum_{\{x, y\}\in E} t_{xy}
 c_x^* c_y \otimes \one \otimes \exp\Big\{
+\im \Phi_{\{x, y\}}
\Big\}.
\end{align} 
Here,  we used that $t_{xy}=t_{yx}$ and $\Phi_{\{y , x\}}=-\Phi_{\{x, y\}}$. 
Thus,  we have (i).
Similarly,  one  obtains that  $\mathcal{W} T_{-g, \downarrow}
	     \mathcal{W}^* =T_{-g, \downarrow}$.

(iii) Since $\mathcal{W} \mathsf{n}_x \otimes \one \mathcal{W}^*=(\one
-\mathsf{n}_x) \otimes \one$ and $\mathcal{W} \one \otimes
\mathsf{n}_x \mathcal{W}^* =\one \otimes \mathsf{n}_x$,
we see that 
\begin{align}
\mathcal{W}\mathbf{U} \mathcal{W}^* =\widetilde{\mathbf{U}}.\ \ \ \ \Box
\end{align} 

\begin{coro}
Let $\mathbb{H}_M=\mathcal{W} \widehat{H}_M
 \mathcal{W}^*$. Then
 we have
\begin{align}
\mathbb{H}_M=-T_{+g, \uparrow}-T_{-g, \downarrow}+\widetilde{\mathbf{U}}+H_{\mathrm{p}}.
\end{align} 
\end{coro} 

\subsection{Expression of the Hamiltonian  in $\mathscr{L}^2(\Fock_{\mathrm{e}, \Md})\otimes L^2(\mathcal{Q})$}

\subsubsection{Natural identification  $\Fock_{\mathrm{e}, \Md}\otimes
   \Fock_{\mathrm{e}, \Md}$ with
   $\mathscr{L}^2(\Fock_{\mathrm{e}, \Md})$ }
Let $\vartheta$ be an anti-linear involution on $\Fock_{\mathrm{e}, \Md}$ defined by
\begin{align}
\vartheta c_x \vartheta=c_x,\ \ \ \vartheta \Omega=\Omega,
\end{align} 
where $\Omega$ is the Fock vacuum in $\Fock_{\mathrm{e}}$.
We define an isometric isomorphism from $\mathscr{L}^2(\Fock_{\mathrm{e}, \Md})$ onto
$\Fock_{\mathrm{e}, \Md}\otimes \Fock_{\mathrm{e}, \Md}$ by 
\begin{align}
\Phi_{\vartheta}\big(|\vphi\ra\la \psi|\big)=\vphi\otimes \vartheta \psi.
\end{align} 
Hence,  we can identify $\mathscr{L}^2(\Fock_{\mathrm{e}, \Md})$ with
$\Fock_{\mathrm{e}, \Md}\otimes \Fock_{\mathrm{e}, \Md}$ by $\Phi_{\vartheta}$.
Moreover,  one has 
\begin{align}
\Phi_{\vartheta}\mathcal{L}(A)\Phi_{\vartheta}^{-1}=A\otimes \one,\ \ \ 
\Phi_{\vartheta}\mathcal{R}(\vartheta A^*\vartheta)\Phi_{\vartheta}^{-1}=\one\otimes A\label{OpIdentify}
\end{align} 
for any bounded linear operator $A$ on $\Fock_{\mathrm{e}, \Md}$.
To summarize, we have the following identifications:
\begin{align}
\Fock_{\mathrm{e}, \Md}\otimes \Fock_{\mathrm{e},
 \Md}=\mathscr{L}^2(\Fock_{\mathrm{e}, \Md}), \label{FockI}
\\
\mathcal{L}(A)=A\otimes \one,\ \ \ 
\mathcal{R}(\vartheta A^*\vartheta)=\one\otimes A.
\end{align} 

\subsubsection{The Schr\"odinger representation}
Note the following identification:
\begin{align}
\mathfrak{P}=L^2(\mathcal{Q}, d {\boldsymbol q})=L^2(\mathcal{Q}),\label{SchI}
\end{align} 
where $\mathcal{Q}=\BbbR^{|\Lambda|}$,  $d {\boldsymbol q}=\prod_{x\in
\Lambda}dq_x$ is the
$|\Lambda|$-dimensional Lebesgue measure on $\mathcal{Q}$, and
$L^2(\mathcal{Q})$ is the Hilbert space of the square integrable
functions on $\mathcal{Q}$.
Under this identification, 
$q_x$ and $p_x$ can be viewed as  multiplication and partial
differential operators, respectively.
Moreover,  the phonon energy term can be expressed as 
\begin{align}
H_{\mathrm{p}}= \frac{1}{2}\sum_{x\in
 \Lambda}\Big(-\nabla_{q_x}^2+\omega_0^2 q^2_x\Big)- \frac{|\Lambda|}{2}
.\label{KPhonon}
\end{align} 
\subsubsection{Representation in $\mathscr{L}^2(\Fock_{\mathrm{e},\Md }) \otimes L^2(\mathcal{Q})$}
By (\ref{FockI}) and (\ref{SchI}), we have the  following identifications:
\begin{align}
\mathscr{L}^2(\Fock_{\mathrm{e}, \Md})\otimes \mathfrak{P}
=\mathscr{L}^2(\Fock_{\mathrm{e},\Md }) \otimes L^2(\mathcal{Q})
=\int^{\oplus}_{\mathcal{Q}} \mathscr{L}^2(\Fock_{\mathrm{e}, \Md})\, d {\boldsymbol
 q}.
\label{SchRep}
\end{align} 
For each $\psi=\int_{\mathcal{Q}}^{\oplus} \psi(\bphi)\, d \bphi\in
\mathscr{L}^2(\Fock_{\mathrm{e}, \Md})\otimes
 L^2(\mathcal{Q})=\int^{\oplus}_{\mathcal{Q}} \mathscr{L}^2(\Fock_{\mathrm{e}, \Md})\, d \bphi$,
let us define an isometric isomorphism $\Phi_{\vartheta}^{\oplus}$
from $\mathscr{L}^2(\Fock_{\mathrm{e}, \Md})\otimes L^2(\mathcal{Q})$ onto 
$[\Fock_{\mathrm{e},\Md }\otimes \Fock_{\mathrm{e}, \Md}]\otimes L^2(\mathcal{Q}) $ by 
\begin{align}
\Phi^{\oplus}_{\vartheta}(\psi)=\int^{\oplus}_{\mathcal{Q}}
\Phi_{\vartheta}(\psi(\bphi))\, d \bphi.
\end{align} 

Let $\bphi\mapsto A(\bphi)$ be a $\mathscr{B}(\Fock_{\mathrm{e}, \Md})$-valued
measurable map such that  $\sup_{\bphi} \|A(\bphi)\|_{\mathscr{B}}<\infty$. Using (\ref{OpIdentify}), we see that 
\begin{align}
\Phi^{\oplus}_{\vartheta}\int^{\oplus}_{\mathcal{Q}}
\mathcal{L}\big(
A(\bphi)
\big)\, d \bphi\,  \Phi_{\vartheta}^{\oplus -1}
&=\int_{\mathcal{Q}}^{\oplus} A(\bphi)\otimes \one\, d \bphi,\\
\Phi^{\oplus}_{\vartheta}\int^{\oplus}_{\mathcal{Q}}
\mathcal{R}\big(
\vartheta A(\bphi)^* \vartheta
\big)\, d \bphi\,  \Phi_{\vartheta}^{\oplus -1}
&=\int_{\mathcal{Q}}^{\oplus}\one \otimes  A(\bphi)\, d \bphi.
\end{align}

\begin{lemm}\label{RepDirect}
Under  identification (\ref{SchRep}), we have the following:
\begin{item}
\item[{\rm (i)}]
\begin{align}
T_{+g, \uparrow}=\int_{\mathcal{Q}}^{\oplus}
 \mathcal{L}(\mathbf{T}_{+g}(\bphi))d\bphi,\ \ \ T_{-g,
 \downarrow}=\int_{\mathcal{Q}}^{\oplus}
 \mathcal{R}(\mathbf{T}_{+g}(\bphi))d\bphi,
\end{align} 
where 
\begin{align}
\mathbf{T}_{\pm g}(\bphi)&=\sum_{\{x, y\}\in E} t_{xy}c_x^* c_y  \exp\Big\{
\pm \im \Phi_{\{x, y\}}(\bphi)
\Big\}
, \label{Tg}\\ \Phi_{\{x, y\}}(\bphi)&=
\sqrt{2}\omega_0^{-1/2}
\sum_{z\in \Lambda} 
(
g_{xz}-g_{yz}
)q_z,
\end{align} 
for each $\bphi=\{q_x\}_x\in \mathcal{Q}$.
\item[{\rm (ii)}]
\begin{align} 
\widetilde{\mathbf{U}}=\frac{1}{2}\sum_{x, y\in \Lambda} U_{\ef, xy}\big\{\mathcal{L}(\mathsf{n}_x)-\mathcal{R}(\mathsf{n}_x)\big\}\big\{\mathcal{L}\big(\mathsf{n}_y)-\mathcal{R}(\mathsf{n}_y)\big\}\otimes\one_{L^2}
,
\end{align} 
 where $\one_{L^2}$ is the identity operator on $L^2(\mathcal{Q})$.
\end{item} 
\end{lemm} 
{\it Proof.} (i) Since $\mathcal{L}(\cdot)$ is linear, i.e.,
$\mathcal{L}(aX+bY) =a \mathcal{L}(X)+b\mathcal{L}(Y)$, we have 
\begin{align}
T_{+g, \uparrow}&=\int_{\mathcal{Q}}^{\oplus}
\sum_{\{x,y\}\in E}t_{xy} \exp\Big\{\im
\Phi_{\{x, y\}}(\bphi)
\Big\}
c_x^* c_y
\otimes \one 
d\bphi\no
&=\int_{\mathcal{Q}}^{\oplus}
\sum_{\{x,y\}\in E}t_{xy} \exp\Big\{\im
\Phi_{\{x, y\}}(\bphi)
\Big\}
\mathcal{L}(c_x^* c_y)
d\bphi\no
&= \int_{\mathcal{Q}}^{\oplus} \mathcal{L}(\mathbf{T}_{+g}(\bphi))d\bphi.
\end{align} 

Similarly, since $\mathcal{R}(\cdot)$ is linear and $\vartheta c_x\vartheta =c_x$,
we have 
\begin{align}
T_{-g, \downarrow}&=\int_{\mathcal{Q}}^{\oplus}
\sum_{\{x,y\}\in E}t_{xy} \exp\Big\{-\im
\Phi_{\{x, y\}}(\bphi)
\Big\}
\one \otimes  c_x^* c_y
d\bphi\no
&=\int_{\mathcal{Q}}^{\oplus}
\sum_{\{x,y\}\in E}t_{xy} \exp\Big\{-\im
\Phi_{\{x, y\}}(\bphi)
\Big\}
\mathcal{R}(\vartheta (c_x^* c_y)^* \vartheta)
d\bphi\no
&=\int_{\mathcal{Q}}^{\oplus}
\sum_{\{x,y\}\in E}t_{xy} \exp\Big\{-\im
\Phi_{\{x, y\}}(\bphi)
\Big\}
\mathcal{R}(c_y^* c_x)
d\bphi\no
&=\int_{\mathcal{Q}}^{\oplus}
\sum_{\{y,x\}\in E}t_{yx} \exp\Big\{-\im
\Phi_{\{y, x\}}(\bphi)
\Big\}
\mathcal{R}(c_x^* c_y)
d\bphi\no
&= \int_{\mathcal{Q}}^{\oplus} \mathcal{R}(\mathbf{T}_{+g}(\bphi))d\bphi.
\end{align} 
Here, we have  used $t_{xy}=t_{yx}$  and $\Phi_{\{y, x\}}(\bphi)=-\Phi_{\{x, y\}}(\bphi)$.

(ii) is immediate. $\Box$

\begin{coro}\label{HMEX}
Under  identification (\ref{SchRep}), we have
\begin{align}
\mathbb{H}_M=-\mathbb{T}-\mathbb{U}+H_{\mathrm{p}},
\end{align} 
where
\begin{align}
\mathbb{T}&=\int_{\mathcal{Q}}^{\oplus}
 \mathcal{L}(\mathbb{T}_{+g}(\bphi))d\bphi
+\int_{\mathcal{Q}}^{\oplus}
 \mathcal{R}(\mathbb{T}_{+g}(\bphi))d\bphi, \label{TFiber}\\
\mathbb{T}_{+g}(\bphi)&= \mathbf{T}_{+g}(\bphi)+\frac{1}{2}\la \mb{n},
 \mathbf{U}_{\mathrm{eff}} \mb{n}\ra,\\
\mathbb{U}&=\sum_{x, y\in \Lambda} U_{\mathrm{eff}, xy}
 \mathcal{L}(\mathsf{n}_x) \mathcal{R}(\mathsf{n}_y) \otimes \one_{L^2}.
\end{align} 
Here,  we use the following notation:
$
\la \mb{n}, \mb{U}_{\mathrm{eff}}\mb{n}\ra:=\sum_{x, y\in \Lambda}
 U_{\mathrm{eff}, xy} \mathsf{n}_x \mathsf{n}_y
$.
\end{coro}

\subsection{Functional integral representation}

Under identification (\ref{SchRep}), each $\psi\in \mathscr{L}^2(\Fock_{\mathrm{e}, \Md})\otimes
L^2(\mathcal{Q})$ can be expressed as 
$\psi=\int_{\mathcal{Q}}^{\oplus } \psi(\bphi)\, d\bphi$, where
$\psi(\bphi) \in \mathscr{L}^2(\Fock_{\mathrm{e}, \Md})$ for a.e. $\bphi$.

\begin{define}
{\rm 
Let $A$ be a bounded linear operator on $\mathscr{L}^2(\Fock_{\mathrm{e},
 \Md})\otimes L^2(\mathcal{Q})$. If there exists a
 $\mathscr{B}(\mathscr{L}^2(\Fock_{\mathrm{e, \Md}}))$-valued map $(\bphi,
 \bphi')\mapsto K(\bphi,
 \bphi')$
 such that 
\begin{align}
(A\psi)(\bphi)=\int_{\mathcal{Q}} K(\bphi, \bphi') \psi(\bphi')d\bphi' \
 \   \mbox{$\forall\psi \in \mathscr{L}^2(\Fock_{\mathrm{e},
 \Md})\otimes L^2(\mathcal{Q})$ },
\end{align} 
then we say that {\it $A$ has a kernel operator $K$. } We denote by $A(\bphi,
 \bphi')$ the kernel
 operator of $A$  if it exists. Trivially,  it holds that 
\begin{align}
\la \vphi, A\psi\ra=\int_{\mathcal{Q}\times \mathcal{Q}} d\bphi d\bphi'
\big \la\vphi(\bphi), A(\bphi, \bphi') \psi(\bphi')
 \big\ra_{\mathscr{L}^2(\Fock_{\mathrm{e}, \Md})}. \ \ \ \diamondsuit
\end{align} 
}
\end{define}   

In this subsection, we will express  the kernel operator  of
$\exp\{-\beta(- \mathbb{T}+H_{\mathrm{p}})\}$ in terms of a functional
integral representation.

In the remainder of this paper, we may assume that $\omega_0=1$ without
loss of generality.

Set $A=C([0, \infty); \mathcal{Q})$, the set of all $\mathcal{Q}$-valued
 continuous functions on $[0, \infty)$.
Let $(A, \mathscr{B}(A), D \alpha)$ be the  probability space for the 
$|\Lambda|$-dimensional Brownian bridge 
 $\{{\boldsymbol\alpha}(s)\, |\, 0\le s
\le 1\}=\{\{\alpha_x(s)\}_{x\in \Lambda}\, |\, 0\le s \le 1\}$, i.e., the Gaussian process with covariance
\begin{align}
\int_A \alpha_x(s)\alpha_y(t)\, D\alpha=\delta_{xy}s(1-t) \label{Bridge}
\end{align}  
for $0\le s \le t \le 1$ and $x, y\in \Lambda$. Define, for each $\bphi, \bphi'\in
\mathcal{Q}$,
\begin{align}
\bomega(s)=(
1-\beta^{-1}s)
\bphi+\beta^{-1}s \bphi'+\sqrt{\beta}{\boldsymbol \alpha}(\beta^{-1}s).\label{DefOmega}
\end{align} 
The conditional Wiener measure $d\mu_{\bphi, \bphi'; \beta}$ is
given by 
\begin{align}
d \mu_{\bphi, \bphi'; \beta}=P_{\beta}(\bphi, \bphi')D\alpha,
\end{align} 
where $P_{\beta}(\bphi,
\bphi')=(2\pi\beta)^{-1/2}\exp\big(-\frac{1}{2\beta}|\bphi-\bphi'|^2\big)$.

For each $\bfphi\in A$, $\bomega(\bfphi)$ indicates a function $s
\mapsto \bomega(s)(\bfphi)$, the sample path $\bomega(\cdot)(\bfphi)$
associated with $\bfphi$.
 Let 
\begin{align}
G_{\beta}(\bomega(\bfphi))=
\prod_0^{{\beta}\atop{\longrightarrow}}
\ex^{\mathbb{T}_{+g}(\bomega(s)(\bfphi))\, d s},\label{GOp}
\end{align} 
where the RHS of (\ref{GOp}) is the strong product integration (see
Appendix \ref{SPI}). Note that since   $\bomega(s)(\bfphi)$ is continuous
in $s$
for all $\bfphi\in A$,  the RHS of (\ref{GOp}) exists.

\begin{Prop}\label{KMkernel}
Let 
\begin{align}
K_M=-\mathbb{T}+H_{\mathrm{p}}.
\end{align} 
Then
$\ex^{-\beta K_M}$ has a kernel operator given by 
\begin{align}
\ex^{-\beta K_M }(\bphi, \bphi')
=&\int d \mu_{\bphi, \bphi'; \beta}\, \mathcal{L}\Big[
G_{\beta}(\bomega)
\Big]
 \mathcal{R}\Big[
G_{\beta}(\bomega)^*\,
\Big]\, 
\ex^{-\int_0^{\beta}d s\,  \mathcal{V}(\bomega(s))}, \label{FKformula}
\end{align} 
   where 
\begin{align}
\mathcal{V}(\bphi)&=\frac{1}{2}\sum_{x\in \Lambda}\omega_0^2q_x^2-\frac{1}{2}|\Lambda|.
\end{align} 
\end{Prop} 
{\it Proof.}
First,  note that 
\begin{align}
&\Big\la f_0, \ex^{-\beta H_{\mathrm{p}}/n}f_1 \ex^{-\beta
 H_{\mathrm{p}}/n}f_2 \cdots f_n\Big\ra\no
=&\int_{\mathcal{Q}\times \mathcal{Q}} d\bphi d\bphi'\int d\mu_{\bphi,
 \bphi'; \beta}\, \ex^{-\int_0^{\beta}ds \mathcal{V}(\bomega(s))}\no
 &\times f_0(\bphi)^* f_1\big(\bomega(\tfrac{\beta}{n})\big)
 f_2\big(\bomega(\tfrac{2\beta}{n})\big) \cdots f_{n-1}\big(\bomega(\tfrac{(n-1)\beta}{n})\big) f_n(\bphi')
\,  \label{FKFormula}
\end{align} 
for  $f_0, f_n\in L^2(\mathcal{Q})$ and $f_1, \dots, f_{n-1}\in
L^{\infty}(\mathcal{Q})$, see \cite{Simon}.
Let $\mathbb{T}(\bphi)=\mathcal{L}(\mathbb{T}_{+g}(\bphi))+\mathcal{R}(\mathbb{T}_{+g}(\bphi))$.
 By (\ref{FKFormula}) and the Trotter--Kato product formula, we have 
\begin{align}
\big\la \vphi, \ex^{-\beta K_M}\psi\big\ra
=&\lim_{n\to \infty} \Big\la \vphi, \Big(
\ex^{- \beta H_{\mathrm{p}}/n} \ex^{\beta \mathbb{T}/n}
\Big)^n \psi\Big\ra\no
=& \lim_{n\to \infty} \int_{\mathcal{Q}\times \mathcal{Q}}d\bphi d\bphi'
\int d\mu_{\bphi, \bphi'; \beta}\, \ex^{-\int_0^{\beta}ds \mathcal{V}(\bomega(s))}\no
&\ \ \  \times \Big\la 
\vphi(\bphi),
 \ex^{\tfrac{\beta}{n}\mathbb{T}(\bomega(\tfrac{\beta}{n}))}
 \ex^{\tfrac{\beta}{n}\mathbb{T}(\bomega(\tfrac{2\beta}{n}))}
\cdots
 \ex^{\tfrac{\beta}{n}\mathbb{T}(\bomega(\tfrac{n \beta}{n}))}
\psi(\bphi')
\Big\ra_{\mathscr{L}^2(\Fock_{\mathrm{e}, \Md})}\no
=& \lim_{n\to \infty} \int_{\mathcal{Q}\times \mathcal{Q}}d\bphi d\bphi'
\int d\mu_{\bphi, \bphi'; \beta}\ex^{-\int_0^{\beta}ds\mathcal{V}(\bomega(s))}\no
&\ \ \ \ \times \bigg\la 
\vphi(\bphi),
 \mathcal{L}
\bigg[
\ex^{\tfrac{\beta}{n}\mathbb{T}_{+g}(\bomega(\tfrac{\beta}{n}))}
\cdots
 \ex^{\tfrac{\beta}{n}\mathbb{T}_{+g}(\bomega(\tfrac{n\beta}{n}))}
\bigg]\no
& \ \ \ \ \times \mathcal{R}
\bigg[
\ex^{\tfrac{\beta}{n}\mathbb{T}_{+g}(\bomega(\tfrac{n\beta}{n}))}
\cdots
\ex^{\tfrac{\beta}{n}\mathbb{T}_{+g}(\bomega(\tfrac{\beta}{n}))}
\bigg]
\psi(\bphi')
\bigg\ra_{\mathscr{L}^2(\Fock_{\mathrm{e}, \Md})}.
\end{align} 
By the dominated convergence theorem, we
conclude (\ref{FKformula}). $\Box$

\section{Proof of Theorem \ref{PositiveGS}} \label{Sec4}
\subsection{Strategy}
The main purpose of this section is to prove  Theorem  \ref{AtLeastOne}
below. As seen  in Subsection \ref{CompletePP}, Theorem
\ref{PositiveGS}
 is a corollary of Theorem \ref{AtLeastOne}.

\begin{Thm}\label{AtLeastOne}
Assume that $|\Lambda|$ is even. Assume {\bf (A. 1)}.  Assume that $\Ue$ is positive semi-definite. 
Then  for all $M\in \{-|\Lambda|/2, -|\Lambda|/2+1, \dots, |\Lambda
|/2 \}$, there exists a Hilbert cone $\mathfrak{H}_{M, +}$ such that 
$
\ex^{-\beta H_M} \unrhd 0$  w.r.t. $ \mathfrak{H}_{M, +}$
holds for all $\beta \ge 0$.
\end{Thm} 

In the remainder of this section, we will continue to assume {\bf
(A. 1)} and 
that 
$|\Lambda|$ is even.

\subsection{Preliminaries}

The canonical cone in  $ \mathscr{L}^2(\Fock_{\mathrm{e},
\Md})\otimes L^2(\mathcal{Q})$  is given by 
\begin{align}
\mathfrak{C}_M=\int^{\oplus}_{\mathcal{Q}} \mathscr{L}^2(\Fock_{\mathrm{e}, \Md})_+d
 \bphi,
\end{align} 
where the direct integral of $\mathscr{L}^2(\Fock_{\mathrm{e}, \Md})_+$ over
$\mathcal{Q}$
 is defined  by
\begin{align}
&\int^{\oplus}_{\mathcal{Q}} \mathscr{L}^2(\Fock_{\mathrm{e}, \Md})_+d\bphi\no
&=\Big\{
\Psi\in \mathscr{L}^2(\Fock_{\mathrm{e}, \Md})\otimes L^2(\mathcal{Q})\, \Big|\,
 \Psi(\bphi) \ge 0\ \mbox{w.r.t. $\mathscr{L}^2(\Fock_{\mathrm{e}, \Md})_+$ for
  a.e. $\bphi$}
\Big\}.
\end{align}
\begin{Prop} 
 $\mathfrak{C}_M$ is a Hilbert cone in $
 \mathscr{L}^2(\Fock_{\mathrm{e},
\Md})\otimes L^2(\mathcal{Q})
$.
\end{Prop} 
{\it Proof.}  We will check the conditions (i)-(iii) of Definition
\ref{HilCone}.

(i) For all $\Phi,\Psi\in \mathfrak{C}_M$, we know that $\la
\Phi(\bphi), \Psi(\bphi)\ra_{\mathscr{L}^2}\ge 0$ for
a.e. $\bphi$. Hence,
$\la \Phi, \Psi\ra=\int_{\mathcal{Q}} \la \Phi(\bphi),
\Psi(\bphi)\ra_{\mathscr{L^2}^2}d\bphi\ge 0$.

(ii) Let $\mathfrak{C}_{M, \BbbR}$ be a real subspace generated by
$\mathfrak{C}_{M}$. It is easy to see that 
$\mathfrak{C}_{M, \BbbR}=\{\Psi\in \mathscr{L}^2(\Fock_{\mathrm{e},
\Md})\otimes L^2(\mathcal{Q})\, |\, \mbox{$\Psi(\bphi)$ is self-adjoint
for a.e. $\bphi$}
\}$. Let $\Psi\in \mathfrak{C}_{M, \BbbR}$. Since
$\mathscr{L}^2(\Fock_{\mathrm{e}, \Md})_+$ is a Hilbert cone, we have a
decomposition $\Psi(\bphi)=\Psi_+(\bphi)-\Psi_-(\bphi)$ such 
$\Psi_{\pm}(\bphi)\in \mathscr{L}^2(\Fock_{\mathrm{e}, \Md})_+ $ 
and $\la \Psi_+(\bphi), \Psi_-(\bphi)\ra_{\mathscr{L}^2}=0$. Thus, (ii)
is clear.

(iii) For each $\Psi\in \mathscr{L}^2(\Fock_{\mathrm{e}, \Md})\otimes
L^2(\mathcal{Q}) $,
 we define $\Psi_{R}, \Psi_I\in \mathfrak{C}_{M, \BbbR}$ by 
$
\Psi_R(\bphi)=\frac{1}{2}(\Psi(\bphi)+\Psi(\bphi)^*),\ 
\Psi_I(\bphi)=\frac{1}{2i}(\Psi(\bphi)-\Psi(\bphi)^*)
$. Then  $\Psi=\Psi_R+i\Psi_I$. $\Box$

\begin{lemm}\label{ConeEquiv}
Let $\Psi\in \mathscr{L}^2(\Fock_{\mathrm{e},
\Md})\otimes L^2(\mathcal{Q})$.
The following are equivalent:
\begin{itemize}
\item[{\rm (i)}] $\Psi\in \mathfrak{C}_M$.
\item[{\rm (ii)}] $\forall \xi\in \mathscr{L}^2(\Fock_{\mathrm{e},
\Md})_+ \forall f\in  L^2(\mathcal{Q})_+,\ \la \Psi, \xi\otimes f\ra \ge
	     0$.
\end{itemize} 
\end{lemm} 
{\it Proof.} To show that (i) $\Rightarrow$ (ii) is easy. 
Let us show the inverse.
Set $g_{\xi}(\bphi)=\la
\Psi(\bphi), \xi\ra_{\mathscr{L}^2}$. By (ii), we have 
\begin{align}
0 \le \la \Psi, \xi\otimes f\ra=\int_{\mathcal{Q}} f(\bphi) g_{\xi}(\bphi)d\bphi.
\end{align} 
From this, we conclude that $g_{\xi}(\bphi)\ge 0$ a.e. $\bphi$. 
Since $\xi$ is
arbitrary, we see that $\Psi\in \mathfrak{C}_{M, \BbbR}$, otherwise, 
$g_{\xi}(\bphi)$ becomes a complex-valued function for some $\xi$.
Since $\mathscr{L}^2(\Fock_{\mathrm{e},
\Md})_+ $ is a Hilbert cone, we have  the decomposition
$\Psi(\bphi)=\Psi_+(\bphi)-\Psi_-(\bphi)$, such that $\Psi_{\pm}(\bphi)
\in \mathscr{L}^2(\Fock_{\mathrm{e},
\Md})_+$ and $\la \Psi_+(\bphi), \Psi_-(\bphi)\ra_{\mathscr{L}^2}=0$.
Since $\xi$ is arbitrary,  by taking $\xi=\Psi_-(\bphi)$, we have 
\begin{align}
0\le g_{\xi}(\bphi)=-\|\Psi_-(\bphi)\|^2\le 0,
\end{align} 
which implies that  $\Psi_-(\bphi)=0$.
Thus, $\Psi\in \mathfrak{C}_M$. $\Box$

\begin{lemm}\label{FiberPP}
Let $B: \mathcal{Q}\to \mathscr{B}(\Fock_{\mathrm{e}, \Md});\,   \bphi \mapsto B(\bphi)$  be strongly continuous.
Then we have 
\begin{align}
\int^{\oplus}_{\mathcal{Q}}
 \mathcal{L}(B(\bphi)^*)\mathcal{R}(B(\bphi))d\bphi\unrhd 0\ \ 
\mbox{w.r.t. $
\mathfrak{C}_M
$}.
\end{align} 
In particular,  $\mathcal{L}(C^*)\mathcal{R}(C)\otimes \one_{L^2} \unrhd 0$
 w.r.t. $\mathfrak{C}_M$ for each $C\in
 \mathscr{B}(\mathscr{L}^2(\Fock_{\mathrm{e}, \Md}))$\footnote{
$\mathscr{B}(\mathscr{L}^2(\Fock_{\mathrm{e}, \Md}))$
 is the set of all bounded linear  operators in the Hilbert space 
$\mathscr{L}^2(
\Fock_{\mathrm{e}, \Md})
$, 
}.
\end{lemm} 
{\it Proof. }
For a.e. $\bphi$, we obtain  $\mathcal{L}(B(\bphi)^*)\mathcal{R}(B(\bphi))\unrhd
0$ w.r.t. $\mathscr{L}^2(\Fock_{\mathrm{e}, \Md})_+$ by Lemma
\ref{GeneralPP}.
Thus,  $
\int^{\oplus}_{\mathcal{Q}}
 \mathcal{L}(B(\bphi)^*)\mathcal{R}(B(\bphi))d\bphi
$ leaves $\mathfrak{C}_M$
invariant. $\Box$
\medskip\\

Let $L^2(\mathcal{Q})_+$ be a  Hilbert cone in
$L^2(\mathcal{Q})$ defined by 
\begin{align}
L^2(\mathcal{Q})_+=\{F\in
L^2(\mathcal{Q})\, |\, F(\bphi)\ge 0\  \mbox{a.e.}\}.
\end{align} 
Then, the following lemma will be useful:
\begin{lemm}\label{TensorL2}
Let $A$ be a bounded linear operator in $L^2(\mathcal{Q})$.
If $A\unrhd 0$ w.r.t. $L^2(\mathcal{Q})_+$, then $\one_{\mathscr{L}^2}
 \otimes A \unrhd 0$ w.r.t. $\mathfrak{C}_M$.
\end{lemm} 
{\it Proof.} Let $f\in L^2(\mathcal{Q})_+$. Since $A\unrhd 0$
w.r.t. $L^2(\mathcal{Q})_+$, we know $Af\in L^2(\mathcal{Q})_+$.
Thus, for each $\xi\in  \mathscr{L}^2(\Fock_{\mathrm{e}, \Md})_+$, 
it holds that $\xi\otimes Af \in \mathfrak{C}_M$. Hence, for each
$\Psi\in \mathfrak{C}_M$, we have
\begin{align}
\la \one_{\mathscr{L}^2} \otimes A \Psi, \xi\otimes f\ra=\la \Psi, \xi \otimes Af\ra\ge 0.
\end{align}
By Lemma \ref{ConeEquiv}, we obtain $\one_{\mathscr{L}^2} \otimes A \Psi\in
\mathfrak{C}_M$, which means that $\one_{\mathscr{L}^2} \otimes A\unrhd 0$
w.r.t. $\mathfrak{C}_M$. $\Box$

\subsection{Lower bounds for the effective  Coulomb interaction}
 
\begin{Prop} \label{CoulombPP}
We have  the following:
\begin{itemize}
\item[{\rm (i)}] If  $\Ue$ is positive semi-definite, then 
\begin{align}
\sum_{x, y\in \Lambda} U_{\ef, xy}\mathcal{L}\big(\mathsf{n}_x\big)\mathcal{R}\big(\mathsf{n}_y\big) \otimes
 \one_{L^2} \unrhd 0 \
 \mbox{w.r.t. $\mathfrak{C}_M$.}
\end{align} 
\item[{\rm (ii)}] If  $\Ue$ is positive definite, then 
	     there exists a $U_0>0$ such that 
\begin{align}
\sum_{x, y\in \Lambda} U_{\ef, xy}\mathcal{L}\big(\mathsf{n}_x\big)\mathcal{R}\big(\mathsf{n}_y\big) \otimes
 \one_{L^2}
 \unrhd U_0\sum_{x\in \Lambda}
 \mathcal{L}(\mathsf{n}_x)\mathcal{R}(\mathsf{n}_x)\otimes \one_{L^2} \unrhd 0\ \ \mbox{w.r.t. $\mathfrak{C}_M$}. 
\end{align} 
\end{itemize} 
\end{Prop} 
{\it Proof.} (i)
Let $\mathbf{M}=(M_{xy})$ be a $|\Lambda|\times |\Lambda|$ matrix defined
by $M_{xy}=U_{\ef, xy}\, (x, y\in \Lambda)$. By  assumption, $\mathbf{M}$ is positive
semi-definite. Thus,  there exists an orthogonal matrix $\mathbf{P}$ such
that $\mathbf{M}=\mathbf{P}\mathbf{D}\mathbf{P}^T$, where $\mathbf{D}=\mathrm{diag}(\lambda_x)$
is a diagonal matrix with $\lambda_x\ge 0$. 
Set $\mathbf{n}=\big\{\mathsf{n}_x\big\}_{x\in \Lambda}$
and set  $\tilde{\mathbf{n}}=\mathbf{P}^T \mathbf{n}$. Denoting
$\tilde{\mathbf{n}}=(\tilde{\mathsf{n}}_x)_{x\in \Lambda}$, we have 
\begin{align}
\sum_{x, y\in \Lambda} U_{\ef, xy}\mathcal{L}\big(\mathsf{n}_x\big)\mathcal{R}\big(\mathsf{n}_y\big)
&=\la \mathcal{L}(\mathbf{n}), \mathbf{M}\mathcal{R}(\mathbf{n})\ra
=\la \mathcal{L}(\tilde{\mathbf{n}}),
 \mathbf{D}\mathcal{R}(\tilde{\mathbf{n}})
\ra\no
&=\sum_{x\in \Lambda} \lambda_x \mathcal{L}(\tilde{\mathsf{n}}_x)
 \mathcal{R}(\tilde{\mathsf{n}}_x). \label{LLCoulomb}
\end{align} 
Clearly,  the RHS of (\ref{LLCoulomb}) is positive  w.r.t.
$\mathfrak{C}_M$
 by Lemma \ref{FiberPP}.

(ii) By assumption, $\mathbf{M}$ is positive definite.
Thus,   the lowest eigenvalue of $\mathbf{M}$  is strictly positive:
$U_0:=\min_x \lambda_x>0$.
Thus,  by (\ref{LLCoulomb}), one sees that 
\begin{align}
\sum_{x, y\in \Lambda}
 U_{\ef, xy}\mathcal{L}\big(\mathsf{n}_x\big)\mathcal{R}\big(\mathsf{n}_y\big)
=&
\sum_{x\in \Lambda} \lambda_x \mathcal{L}(\tilde{\mathsf{n}}_x)
 \mathcal{R}(\tilde{\mathsf{n}}_x)\no
\unrhd& 
U_0\sum_{x\in
 \Lambda}\mathcal{L}(\tilde{\mathsf{n}}_x)\mathcal{R}(\tilde{\mathsf{n}}_x)\no
=&U_0 \sum_{x\in \Lambda}
 \mathcal{L}\big(\mathsf{n}_x\big)\mathcal{R}\big(\mathsf{n}_x\big)\no
\unrhd &0\ \ \mbox{w.r.t. $\mathscr{L}^2(\Fock_{\mathrm{e}, \Md})_+$}.
\end{align} 
By Lemma \ref{FiberPP}, we conclude our proof of  (ii). $\Box$

\subsection{Completion of proof of Theorem \ref{AtLeastOne}}

\begin{Prop}\label{HamiltonianPP}
Assume that $\Ue$ is positive semi-definite. For all  $\beta \ge 0$ and  $\Md\in
 \{-|\Lambda|/2, -|\Lambda|/2+1, \dots, |\Lambda|/2\}$, we have
$
\ex^{-\beta  \mathbb{H}_M} \unrhd 0
$
w.r.t. $\mathfrak{C}_M$.
\end{Prop} 
{\it Proof.}
Since $\mathbb{U}\unrhd 0$ w.r.t. $\mathfrak{C}_M$ by
Proposition \ref{CoulombPP}, we
have 
\begin{align}
\ex^{\beta
 \mathbb{U}}=\sum_{n=0}^{\infty}\underbrace{\frac{\beta^n}{n!}}_{\ge 0} 
\underbrace{\mathbb{U}^n}_{\unrhd 0}
 \unrhd 0\ \   \mbox{w.r.t. $\mathfrak{C}_M$ for all
 $\beta \ge 0$}.
\end{align} 
By (\ref{KPhonon}) and Lemma \ref{TensorL2}, it holds that $\ex^{-\beta H_{\mathrm{p}}}\rhd 0$
w.r.t. $\mathfrak{C}_M$ for all  $\beta \ge 0$\footnote{
To be precise,  we know that 
$\exp\{-\beta
\frac{1}{2}\sum_x( -\nabla^2_{q_x}+\omega_0^2 q_x^2
)
 \} \unrhd 0$ w.r.t. $L^2(\mathcal{Q})_+$.
Thus,  by Lemma \ref{TensorL2}, we have 
$
\ex^{-\beta H_{\mathrm{p}}}
=\one_{\mathscr{L}^2}\otimes  \exp\{-\beta
\frac{1}{2}\sum_x( -\nabla^2_{q_x}+\omega_0^2 q_x^2
)
 \} \unrhd 0
$ 
w.r.t. $\mathfrak{C}_M$.
}.
Denoting  $K=H_{\mathrm{p}}- \mathbb{U}$, we have  
$\ex^{-\beta K}=\ex^{-\beta H_{\mathrm{p}}} \ex^{\beta \mathbb{U}}
\unrhd 0$ w.r.t. $\mathfrak{C}_M$ for all $\beta \ge 0$.

By (\ref{TFiber}) and Lemma \ref{FiberPP}, we have 
\begin{align}
\ex^{\beta \mathbb{T}}
=\int_{\mathcal{Q}}^{\oplus} \mathcal{L}
\Big(
\ex^{\beta \mathbb{T}_{+g}(\bphi)}
\Big)
\mathcal{R}
\Big(
\ex^{\beta \mathbb{T}_{+g}(\bphi)}
\Big)
d\bphi \unrhd 0\ \  \mbox{w.r.t. $\mathfrak{C}_M$}.
\end{align} 
Combining these properties, we obtain 
\begin{align}
\Big(
\underbrace{
\ex^{\beta \mathbb{T}/n}
}_{\unrhd 0}
\underbrace{
 \ex^{-\beta K/n}
}_{\unrhd 0}
\Big)^n \unrhd 0 \ \  \mbox{w.r.t. $\mathfrak{C}_M$ for all $\beta \ge 0$}.
\end{align} 
Thus,  the proposition follows from the Trotter--Kato formula.
 $\Box$

\subsection{Proof of Theorem  \ref{PositiveGS}}\label{CompletePP}
Let $J$ be a conjugation defined by $(J\Psi)(\bphi)=\Psi^*(\bphi)$
for each $\Psi\in \mathscr{L}^2(\Fock_{\mathrm{e}, \Md})\otimes L^2(\mathcal{Q})$.
Since $\ex^{-\beta \mathbb{H}_M}$ preserves the positivity
w.r.t. $\mathfrak{C}_M$, $\mathbb{H}_M$ commutes with $J$. Let
$\lambda$
 be an eigenvalue of $\mathbb{H}_M$ and let $\Psi$ be a corresponding
 eigenvector. Set $\Psi_{\mathrm{R}}=(\Psi+J\Psi)/2$ and
 $\Psi_{\mathrm{I}}=(\Psi-J\Psi)/2\im $. 
Then $\Psi_{\mathrm{R}}(\bphi)$ and $\Psi_{\mathrm{I}}(\bphi)$
are self-adjoint for a.e. $\bphi$. In addition, they are eigenvectors of
$\mathbb{H}_M$ with  an  associated eigenvalue $\lambda$.

Let $\psi_M$ be a ground state of $\mathbb{H}_M$. 
$\psi_M$ can  be written as 
$\psi_M=\int^{\oplus}_{\mathcal{Q}}
\psi_M(\bphi)d\bphi$ under  identification (\ref{SchRep}).
By the observation above, we may assume that 
$\psi_M(\bphi)$ is self-adjoint for a.e. $\bphi$ without loss of generality. 
Let $\psi_{M, +}(\bphi)$ (resp. $\psi_{M, -}(\bphi)$)
be the positive (resp. negative) part of $\psi_{M}(\bphi)$\footnote{
Precise definitions of $\psi_{M, \pm}(\bphi)$ are given in the proof of
Proposition \ref{SDL2}.

}.
Hence, it holds that  
 $\psi_{M}=\psi_{M, +}-\psi_{M, -},  \psi_{M, \pm}\in  \mathfrak{C}_M$ and $\la \psi_{M, +},
\psi_{M, -}\ra=0$.
By Proposition  \ref{HamiltonianPP}, we have 
\begin{align}
\ex^{-\beta E_M} = \la \psi_{M}, \ex^{-\beta \mathbb{H}_M}
 \psi_{M}\ra \le \la |\psi_{M}|, \ex^{-\beta
 \mathbb{H}_M}|\psi_{M}|\ra, \label{PPInqE}
\end{align} 
where $|\psi_{M}|=\psi_{M, +}+ \psi_{M, -}$.
This means that $|\psi_{M}|$ is a ground state of $\mathbb{H}_M$ as
well.  We will show that $|\psi_M|$ satisfies properties (i) and (ii) in
Theorem \ref{PositiveGS}.

Using the notation in Subsection  \ref{Graphetc}, we can express
$|\psi_M|$
as 
\begin{align}
|\psi_M|=\sum_{x, Y\in
 \wedge^{\Md}\Lambda}\int^{\oplus}_{\mathcal{Q}}|\psi_M|_{XY}(\bphi)|e_X\ra\la 
 e_{Y}| d\bphi.
\end{align}
 Since $\psi_M$ is a non-zero vector, $|\psi_M|$ is non-zero as
 well. Thus,  there exists an $X_0\in \wedge^{\Md}\Lambda$ and a
 measurable
set $\mathcal{I}\subseteq \mathcal{Q}$ with $|\mathcal{I}|>0$
 such that 
$|\psi_M|_{X_0X_0}(\bphi)\neq 0$ for all $\bphi\in \mathcal{I}$ ($X_0$ may depend
 on $\bphi$)\footnote{
Since $|\psi_M|$ is non-zero, there exists a measurable set $\mathcal{I}$ with
 $|\mathcal{I}|>0$
 such that $|\psi_M|(\bphi)\neq 0$  for all $\bphi\in \mathcal{I}$.
For each $\bphi\in \mathcal{I}$, we observe that 
\begin{align}
0<\Tr[|\psi_M|(\bphi)]=\sum_{X\in \wedge^{\Md} \Lambda} |\psi_M|_{XX}(\bphi).
\end{align} 
Hence, there exists an $X_0\in \wedge^{\Md}\Lambda$ such that
$|\psi_M|_{X_0X_0}(\bphi) \neq 0$.
}. Observe that $S^2_{\mathrm{tot}}
 |e_{X_0}\ra\la e_{X_0}|=0$.
From this,  it follows that $\mathsf{P}_{S=0} \psi_M\neq 0$, where $\mathsf{P}_{S=0}$ is
the orthogonal projection onto $\ker[S_{\mathrm{tot}}^2]$.
 Using the fact that $\mathcal{W}^* S_{\mathrm{tot}}^2
 \mathcal{W}=\tilde{S}^2$, we obtain (i).

Let $\vphi_M$ be a positive ground state of $H_M$ and let
$\tilde{\vphi}_M$ be its representation in $\mathscr{L}^2(\Fock_{\mathrm{e},
\Md}) \otimes L^2(\mathcal{Q})$.
Note that $\mathcal{W}
S_{x+}S_{y-}\mathcal{W}^*=\gamma_x\gamma_y\mathcal{L}(c_xc_y^*)
\mathcal{R}((c_xc_y^*)^*)$.
Hence,  
\begin{align}
\big\la \vphi_M, S_{x+}S_{y-}\vphi_M\big\ra=\gamma_x\gamma_y \big\la
 \tilde{\vphi}_M, \mathcal{L}(c_xc_y^*)
 \mathcal{R}((c_xc_y^*)^*)\tilde{\vphi}_M\big\ra. \label{InnerSpin}
\end{align}
Since  $\tilde{\vphi}_M$ is  positive and  $\mathcal{L}(c_xc_y^*)
\mathcal{R}((c_xc_y^*)^*) \unrhd 0$ w.r.t. $\mathfrak{C}_M$,
we conclude  our proof of (ii).
$\Box$

\section{Proof of Theorem  \ref{Uniqueness}} \label{Sec5}

\subsection{Strategy}\label{PPSt}\label{StrategyPI}

Our main purpose in this section is to show Theorem \ref{PI} below. To
this end, 
recall the expression of $\mathbb{H}_M$ in Corollary \ref{HMEX}.
\begin{Thm}\label{PI}
Assume that $|\Lambda|$ is even.  Assume {\bf (A. 1)} and {\bf (A. 2)}.
Assume that  $\Ue$ is positive definite.
For all  $\beta>0$ and $\Md\in \{-|\Lambda|/2, -|\Lambda|/2+1, \dots,
 |\Lambda|/2\}$, we have 
$
\ex^{-\beta \mathbb{H}_M }\rhd 0
$
w.r.t. $\mathfrak{C}_M$.
\end{Thm} 

As a corollary, we obtain the following  result  by Theorem \ref{Faris}.

\begin{coro}\label{SPositive}
Assume that $|\Lambda|$ is even.  Assume {\bf (A. 1)} and {\bf (A. 2)}.
Assume that $\Ue$ is positive definite. Let $E_M$ be the ground state
 energy, i.e., the lowest eigenvalue
 of  $\mathbb{H}_M$.  For each  $\Md\in \{-|\Lambda|/2, -|\Lambda|/2+1, \dots, |\Lambda|/2\}$,
   $E_M$ is nondegenerate and the corresponding
 eigenvector is strictly positive 
 w.r.t. $\mathfrak{C}_M$.
\end{coro}

By this result, we see  the uniqueness claimed in Theorem
\ref{Uniqueness}.
Some additional observations tell us more detailed information about the
ground state stated in Theorem \ref{Uniqueness}; see Subsection
\ref{DetailedGS}.

In the remainder of this section, we continue to make  every assumption
named  in Theorem
\ref{PI}.

Now, let us  explain how to prove  Theorem \ref{PI}.
\begin{Prop}\label{Reduction}
Let $U_0$ be a  strictly  positive constant given by Proposition
 \ref{CoulombPP}. Let 
\begin{align}
\mathbb{U}_0=U_0\sum_{x\in \Lambda}
 \mathcal{L}(\mathsf{n}_x)\mathcal{R}(\mathsf{n}_x)\otimes \one_{L^2}.
\end{align} 
We define a new Hamiltonian $\mathbb{H}_M^{(0)}$ by 
\begin{align}
\mathbb{H}_M^{(0)}=K_M-\mathbb{U}_0.
\end{align} 
If $\ex^{-\beta \mathbb{H}_M^{(0)}} \rhd 0$ w.r.t. $\mathfrak{C}_M$ for
 all $\beta >0$, then $\ex^{-\beta \mathbb{H}_M}\rhd 0$
 w.r.t. $\mathfrak{C}_M$ for all $\beta>0$.
\end{Prop} 
{\it Proof.} By
Proposition \ref{CoulombPP}, it holds that $\mathbb{U}\unrhd
\mathbb{U}_0$ w.r.t. $\mathfrak{C}_M$. Hence,  by applying Proposition
\ref{Mono}, we have $\ex^{-\beta \mathbb{H}_M} \unrhd \ex^{-\beta
\mathbb{H}_M^{(0)}}$ w.r.t. $\mathfrak{C}_M$. Thus,  if $\ex^{-\beta
\mathbb{H}_M^{(0)}}\rhd 0$ w.r.t. $\mathfrak{C}_M$, we conclude that
$\ex^{-\beta \mathbb{H}_M} \rhd 0$ w.r.t. $\mathfrak{C}_M$. $\Box$
\medskip\\

By Proposition \ref{Reduction}, it is sufficient to prove that $\ex^{-\beta
\mathbb{H}_M^{(0)}} \rhd 0$ w.r.t. $\mathfrak{C}_M$ for all $\beta >0$.

By the Duhamel formula, we have the following norm-convergent expansion:
\begin{align}
\ex^{-\beta  \mathbb{H}_M^{(0)}}&=\sum_{n\ge 0} \mathscr{D}_{n, \beta}
,\label{Duhamel2}\\ 
\mathscr{D}_{n, \beta}&=\int_{S_n(\beta)}\ex^{-s_1 K_M} \mathbb{U}_0\, 
\ex^{-s_2 K_M}\mathbb{U}_0\cdots \ex^{-s_n  K_M} \mathbb{U}_0\, 
\ex^{-(\beta-\sum_{j=1}^n s_j)K_M}, \label{Duhamel2}
\end{align} 
 where $\int_{S_n(\beta)}=\int_0^{\beta}d t_1 \int_0^{\beta-t_1}d
t_2\cdots\int_0^{\beta-\sum_{j=1}^{n-1}t_j} d t_n
$ and $\mathscr{D}_{0, \beta}=\ex^{-\beta K_M}$.
In Subsection \ref{ProofErg}, we will prove the following:

\begin{Thm}\label{Ergodicity}(Ergodicity)
$\{\mathscr{D}_{n, \beta}\}_{n\in \BbbN_0}$ is ergodic in the  sense that 
 for each $\vphi, \psi\in \mathfrak{C}_M \backslash \{0\}$,
there are  $\beta>0$ and $n\in \BbbN_0:=\{0\} \cup \BbbN$ such  that
$\la \vphi, \mathscr{D}_{n, \beta}\psi\ra>0$. 
\end{Thm} 

Assuming Theorem \ref{Ergodicity}, we can  prove Theorem \ref{PI}.
\begin{flushleft}
{\it  Proof of Theorem \ref{PI} given Theorem \ref{Ergodicity}}
\end{flushleft} 
The basic idea originates  from \cite{JFroehlich1, Miyao2}.
Note that since $\ex^{\beta \mathbb{T}}\unrhd 0$ and $ \mathbb{U}_0\unrhd 0$
w.r.t. $\mathfrak{C}_M$, we see  that $\mathscr{D}_{n, \beta}\unrhd
0$ w.r.t. $\mathfrak{C}_M$. Thus,  for each  $n\in \BbbN_0$,
one has 
\begin{align}
\ex^{-\beta \mathbb{H}_M^{(0)}}\unrhd \mathscr{D}_{n, \beta} \label{LowerBound}
\end{align} 
w.r.t. $\mathfrak{C}_M$.
Take $\vphi, \psi\in \mathfrak{C}_M\backslash \{0\}$
arbitrarily.
Then by Theorem \ref{Ergodicity}, there exist $\beta>0$ and $n\in
\BbbN_0$ such that $\la \vphi, \mathscr{D}_{n, \beta}\psi\ra>0$.
Hence,  using (\ref{LowerBound}), we have
$
\la \vphi, \ex^{-\beta \mathbb{H}_M^{(0)}}\psi\ra \ge \la \vphi,
 \mathscr{D}_{n, \beta}\psi\ra>0. 
$
To summarize, for each  $\vphi, \psi\in \mathfrak{C}_M\backslash \{0\}$,
there exists a $\beta>0$ such   that 
$\la \vphi, \ex^{-\beta \mathbb{H}_M^{(0)}}\psi\ra>0$.
This means that $\ex^{-\beta \mathbb{H}_M^{(0)}}$ improves the positivity
w.r.t. $\mathfrak{C}_M$,  according to  Theorem \ref{Faris}.
$\Box$
\medskip

\begin{flushleft}
{\bf Conclusion:}\\
It suffices to show Theorem \ref{Ergodicity} to prove Theorem
 \ref{Uniqueness}.
$\diamondsuit$
\end{flushleft} 

\subsection{Preliminaries}\label{Graphetc}
Before we enter the proof of Theorem \ref{Ergodicity}, we need to make some
preparations.

Let $G=(\Lambda, E)$ be a connected graph.
For each $0\le n \le |\Lambda|$,
we set 
\begin{align}
 \Lambda^{(n)}=\big\{
X=(x_1,\dots, x_n)\in \Lambda^n\, \big|\, x_1\neq\cdots\neq x_n
\big\}.
\end{align} 
Let $\mathfrak{S}_n$ be the permutation group on the set $\{1, \dots,
n\}$. Let $(x_1, \dots, x_n), \\(y_1, \dots, y_n)\in \Lambda^{(n)}$. If
there exists a $\sigma \in \mathfrak{S}_n$ such that $(x_{\sigma(1)},
\dots, x_{\sigma(n)})=(y_1, \dots, y_n)$, then we write   $(x_1, \dots,
x_n)\sim (y_1, \dots, y_n)$. The binary relation ``$\sim$'' on
$\Lambda^{(n)}$
 is an equivalence relation.  We denote by $\wedge^{n}\Lambda$ the
 quotient set $\Lambda^{(n)} \backslash \sim$. For notational simplicity,
we  denote by  $(x_1, \dots,
x_n)$
the equivalence class $[(x_1, \dots, x_n)]$ if no confusion occurs. 
We say that $X=(x_1,\dots, x_n),
 Y=(y_1, \dots, y_n)\in \wedge^n \Lambda$ are
 {\it neighbors} if there exists
a unique  $j$ such  that $x_j$ and $y_j$ are neighbors in $G$\footnote{
$x, y\in \Lambda$ is said to be {\it neighbors} if $\{x, y\}\in E$.
}  and  $x_i=y_i$
holds for
all $i\in \{1, \dots, n\}\backslash \{j\}$.
For each $n\in \BbbN_0$, we define a graph $\wedge^n G$
by
\begin{align}
\wedge^n G&=(\wedge^n \Lambda, \wedge^n E),\\
\wedge^n E&=\big\{
\{X, Y\}\in [\wedge^n \Lambda]^2\, |\,  \mbox{$X, Y$ are neighbors}
\big\}
\end{align} 
with $\wedge^0 G=(\emptyset, \emptyset)$, the empty graph, and
$\wedge^1 G=G$.
Remark that since $|\wedge^{|\Lambda|} \Lambda|=1$, $\wedge^{|V|} G$ is trivial.

The following proposition is often useful:

\begin{Prop}\label{ConnectFermi}
If $G$ is connected, then $\wedge^n G$ is connected for all $0<
 n < |\Lambda|$.
\end{Prop}
{\it Proof.} See  \cite{FL, Miyao2}. $\Box$
\medskip

A {\it path} in $\wedge^n G$ is a graph $P=(v, e)\subseteq \wedge^n G$
with $v=\{X_1, \dots, X_N\}$ and\\ $e=\{\{X_1, X_2\},\{X_2, X_3\}, \dots,
\{X_{N-1}, X_N\}\}$, where all $X_j$ are distinct.
The path $P$ is simply denoted by $P=X_1 X_2\cdots X_N$. 
The number $N-1$ is called {\it the  length of  path $P$} and denoted by
$|P|$.
For each $X, Y\in \wedge^n \Lambda$, we denote
by $\mathscr{P}_{XY}^{(n)}$
 the set of all paths from
$X$ to $Y$. For each $L\in \BbbN$, we set 
\begin{align}
\mathscr{P}_{XY}^{(n)}[L]=\Big\{
P\in \mathscr{P}_{XY}^{(n)}\, \Big|\, |P|=L
\Big\}.
\end{align} 
Clearly,  it holds that $\mathscr{P}^{(n)}_{XY}=\bigcup_L
\mathscr{P}_{XY}^{(n)}[L]$.

Let $e_x(y)=\delta_{xy}$. Then $\{e_x\, |\, x\in \Lambda\}$ is a
complete orthonormal system(CONS) of $\ell^2(\Lambda)$.
For each $X=(x_1, \dots, x_n)\in \wedge^{n} \Lambda$, we define 
\begin{align}
e_X=e_{x_1}\wedge\cdots \wedge e_{x_n} \in \wedge^{n}\ell^2(\Lambda).
\end{align} 
Then $\{e_X\, |\, X\in \wedge^{n}\Lambda\}$ is a CONS of
$\wedge^{n}\ell^2(\Lambda)$ as well. Note that each $\psi\in
\mathscr{L}^2(\Fock_{\mathrm{e}, \Md})\otimes L^2(\mathcal{Q})$ can be expressed
as 
\begin{align}
\psi=\sum_{X, Y\in \wedge^{\Md}\Lambda}\int^{\oplus}_{\mathcal{Q}}
 \psi_{XY}(\bphi) |e_X\ra\la e_Y| d\bphi. \label{L2CONSEX}
\end{align}

\subsection{Proof of Theorem \ref{Ergodicity}}\label{ProofErg}

We will prove Theorem \ref{Ergodicity}  step-by-step.
\begin{Prop}
Let 
\begin{align}
\mathscr{C}_{n, \beta}&=
\Big(
\mathbb{U}_0^{\Md}\ex^{-\beta K_M/(n-1)}
\Big)^{n-1} \mathbb{U}_0^{\Md}\no
&=\mathbb{U}^{\Md}_0\ex^{\beta K_M/(n-1)}
 \mathbb{U}^{\Md}_0\cdots \ex^{\beta K_M/(n-1)}
 \mathbb{U}^{\Md}_0. \label{Integrand2}
\end{align} 
Suppose that  $\{\mathscr{C}_{n, \beta}\}$ is ergodic in the  sense
 that, 
 for each  $\vphi, \psi\in \mathfrak{C}_M \backslash \{0\}$,
there exist  $\beta>0$ and $n\in \BbbN_0$ such  that
$\la \vphi, \mathscr{C}_{n, \beta}\psi\ra>0$. Then $\{\mathscr{D}_{n,
 \beta}\}$ is ergodic.
\end{Prop} 
{\it Proof.} 
Set $N(n)=n\Md+(n-1)$.
It suffices to  show that a subsequence $\{\mathscr{D}_{N(n),
\beta}\}_{n, \beta}$ is ergodic.
Let 
\begin{align}
F_n(s_1, \dots, s_n)=\ex^{-s_1 K_M} \mathbb{U}_0\, 
\ex^{-s_2 K_M}\mathbb{U}_0\cdots \ex^{-s_n K_M} \mathbb{U}_0\, 
\ex^{-(\beta-\sum_{j=1}^n s_j)K_M}.
\end{align} 
By (\ref{Duhamel2}), it holds that 
\begin{align}
\mathscr{D}_{N(n), \beta}=\int_{S_{N(n)}(\beta)}F_{N(n)}(s_1, \dots, s_{N(n)}).
\end{align}  
Remark that 
\begin{align}
\mathscr{C}_{n,\beta}=F_{N(n)}\Big(\underbrace{0, \dots,
 0}_{\Md}, \beta/(n-1), \underbrace{0, \dots,
 0}_{\Md}, \dots, \beta/(n-1), \underbrace{0, \dots,
 0}_{\Md}
\Big). \label{DefC}
\end{align} 
In particular, $\mathscr{C}_{n, \beta}\unrhd 0$ w.r.t. $\mathfrak{C}_M$
for all $n\in \BbbN_0$ and $\beta \ge 0$.
Since $\{\mathscr{C}_{n, \beta}\}$ is ergodic,  for each $\vphi, \psi\in \mathfrak{C}_M \backslash \{0\}$,
there are  $\beta>0$ and $n\in \BbbN_0$ such  that
$\la \vphi, \mathscr{C}_{n, \beta}\psi\ra>0$. 
Let $f(s_1,\dots, s_{N(n)})=\la \vphi, F_{N(n)}(s_1,\dots,
s_{N(n)})\psi\ra$.
Then $f$ is a  non-zero positive function such that 
\begin{align}
f\Big(\underbrace{0, \dots,
 0}_{\Md}, \beta/(n-1), \underbrace{0, \dots,
 0}_{\Md}, \dots, \beta/(n-1), \underbrace{0, \dots,
 0}_{\Md}
\Big)>0
\end{align} 
 by (\ref{DefC}).
Moreover,  $f$ is continuous in
$s_1, \dots, s_{N(n)}$. Thus,
\begin{align}
\la \vphi, \mathscr{D}_{N(n), \beta}\psi\ra=\int_{S_{N(n)}(\beta)}
 f(s_1,\dots, s_{N(n)})>0.
\end{align}  
This means that $\{\mathscr{D}_{N(n), \beta}\}_{n, \beta}$ is ergodic. $\Box$
\medskip

In the remainder of this subsection, we will prove that
$\{\mathscr{C}_{n, \beta}\}$ is ergodic.
Henceforth, we may assume  that 
\begin{align}
U_0=1
\end{align} 
without loss of generality.
Let $\wedge^n G=(\wedge^n \Lambda, \wedge^n E)$ be the graph defined in
Subsection \ref{Graphetc}.
\begin{lemm}\label{LowerCl}
Let $E_X=|e_X\ra\la e_X|$ for each  $X\in \wedge^{\Md}\Lambda$.
We have  
\begin{align}
\mathbb{U}^{\Md}_0\unrhd  \sum_{X\in
 \wedge^{\Md}\Lambda}\mathcal{L}(E_X)\mathcal{R}(E_X)\otimes \one_{L^2}\ \ \
 \mbox{w.r.t. $\mathfrak{C}_M$}. \label{Coulomb}
\end{align} 
\end{lemm} 
{\it Proof.}
Since $|\Lambda^{\Md}| \ge |\wedge^{\Md} \Lambda|$ 
and $E_X=\mathsf{n}_{x_1}\dots \mathsf{n}_{x_{\Md}}$ for each $X=(x_1,\dots, x_{\Md})\in
\wedge^{\Md} \Lambda$, we obtain, by Lemma \ref{FiberPP},  
\begin{align}
\mathbb{U}_0^{\Md}&=\sum_{(x_1,\dots, x_{\Md})\in \Lambda^{\Md}}\mathcal{L}\big(
\mathsf{n}_{x_1}\cdots \mathsf{n}_{x_{\Md}}
\big)
\mathcal{R}\big(
\mathsf{n}_{x_1}\cdots \mathsf{n}_{x_{\Md}}
\big)\otimes \one_{L^2}\no
&\unrhd \sum_{X\in
 \wedge^{\Md}\Lambda}\mathcal{L}(E_X)\mathcal{R}(E_X)\otimes \one_{L^2}
\end{align} 
w.r.t. $\mathfrak{C}_M$. $\Box$
\medskip

 We introduce the following notation:
\begin{align}
&\int d \nu_{\bphi, \bphi'; \beta}^{(n-1)}
F(\bomega_1, \dots, \bomega_{n-1})\no
:=&\int_{\mathcal{Q}^{n-2}} \prod_{j=1}^{n-2} d \bphi_j \int d \mu_{\bphi, \bphi_1;
 \beta}({\boldsymbol \vphi}_1)\, d \mu_{\bphi_1, \bphi_2;
 \beta}({\boldsymbol \vphi}_2)
\cdots 
d\mu_{\bphi_{n-2}, \bphi';
 \beta}({\boldsymbol \vphi}_{n-1})\no
&\times \exp\Bigg[
-\sum_{j=1}^{n-1} \int_0^{\beta} d s \mathcal{V}\big(\bomega_j(s)(\boldsymbol \vphi_j)\big)
\Bigg] F\Big(\bomega_1({\boldsymbol \vphi_1}), \dots,
 \bomega_{n-1}(\boldsymbol \vphi_{n-1})\Big). \label{DefNot}
\end{align} 

\begin{rem}
{\rm
Using the Brownian bridge ${\boldsymbol \alpha}_j\, (j=1, \dots, n)$,
 $\bomega_j$ can be expressed as 
\begin{align}
\bomega_j(s)(\bfphi_j)=(1-\beta^{-1}s)\bphi_{j-1}+\beta^{-1}s \bphi_j+
\sqrt{\beta} {\boldsymbol \alpha}_j(\beta^{-1}s)(\bfphi_j). \ \ \
 \diamondsuit
\label{AlpahOmega}
\end{align} 
}
\end{rem} 

\begin{Prop}
For each $P=X_1 X_2\cdots X_{|P|+1}\in \mathscr{P}_{XY}^{(\Md)}$ and
 ${\boldsymbol \vphi}_1, \dots, {\boldsymbol \vphi}_{|P|}\in A$, let 
\begin{align}
\mathrm{\mathscr{G}}_{\beta}^{(\Md)}
\Big(P, \{\bomega_j({\boldsymbol \vphi}_j)\}_{j=1}^{|P|}\Big)
= \prod _{j=1}^{{|P|}\atop{\longrightarrow} } E_{X_j}G_{\beta}(\bomega_j(\bfphi_j))E_{X_{j+1}},
\end{align}
where $\displaystyle\prod_{j=1}^{{n}\atop{\longrightarrow}}A_j:=A_1A_2\cdots
 A_n$, the ordered product.  Set $\tilde{\beta}=\beta/(n-1)$.
 The kernel operator  of  $\mathscr{C}_{n, \beta}$ satisfies the following operator
 inequality : 
\begin{align}
\mathscr{C}_{n, \beta}(\bphi, \bphi')
\unrhd&\sum_{X_1, X_{n}\in \wedge^{\Md}\Lambda}
\sum_{P\in \mathscr{P}_{X_1X_{n}}^{(\Md)}[n-1]}
\int d \nu_{\bphi, \bphi'; \tilde{\beta}}^{(n-1)}\no
&\times\mathcal{L}\bigg[
 \mathrm{\mathscr{G}}_{
\tilde{\beta}}^{(\Md)} \Big(P, \{\bomega_j\}_{j=1}^{n-1}\Big)
\bigg]
\mathcal{R}\bigg[\Big\{
 \mathrm{\mathscr{G}}_{
\tilde{\beta}}^{(\Md)}\Big(P, \{\bomega_j\}_{j=1}^{n-1}\Big)
\Big\}^*
\bigg]
  \label{InqCKernel2}
\end{align} 
w.r.t. $\mathscr{L}^2(\Fock_{\mathrm{e}, \Md})_+$.
\end{Prop}  
{\it Proof.}
First, we note the following fact: Let $A, B$ be bounded operators on
$\mathscr{L}^2(\Fock_{\mathrm{e}, \Md})\otimes
L^2(\mathcal{Q})$. Suppose that $A$ and $B$ have kernel operators.
If $A\unrhd B$ w.r.t. $\mathfrak{C}_M$, then $A(\bphi, \bphi')\unrhd
B(\bphi, \bphi')$ w.r.t. $\mathscr{L}^2(\Fock_{\mathrm{e}, \Md})_+$ for
a.e. $\bphi, \bphi'$\footnote{
The proof of this fact is as follows. Since $A\unrhd B$
w.r.t. $\mathfrak{C}_M$, we have 
$
\la \vphi\otimes f, A\psi\otimes g\ra \ge \la \vphi\otimes f, B\psi\otimes g\ra
$
for all $f, g\in L^2(\mathcal{Q})_+$ and $\vphi, \psi\in
\mathscr{L}^2(\Fock_{\mathrm{e}, \Md})_+$. This means that 
$
\int f(\bphi)g(\bphi')\la \vphi, A(\bphi, \bphi')\psi\ra d\bphi d\bphi'
\ge 
\int f(\bphi)g(\bphi')\la \vphi, B(\bphi, \bphi')\psi\ra d\bphi d\bphi'
$. Thus, $
\la \vphi, A(\bphi, \bphi')\psi\ra
\ge 
\la \vphi, B(\bphi, \bphi')\psi\ra
$
holds for a.e. $\bphi, \bphi'$.
Since $\vphi, \psi\in \mathscr{L}^2(\Fock_{\mathrm{e}, \Md})_+$, we
conlude that 
 $A(\bphi, \bphi')\unrhd
B(\bphi, \bphi')$ w.r.t. $\mathscr{L}^2(\Fock_{\mathrm{e}, \Md})_+$ for
a.e. $\bphi, \bphi'$.
}.

By Proposition \ref{KMkernel} and  Lemma \ref{LowerCl}, we have   
\begin{align}
&\mathscr{C}_{n, \beta}(\bphi, \bphi')\no
\unrhd& \sum_{X_1,\dots, X_{n}\in
 \wedge^{\Md} \Lambda} 
\int d \nu_{\bphi, \bphi'; \tilde{\beta}}^{(n-1)}\no
&\times \mathcal{L}
\bigg[
E_{X_1}
G_{\tilde{\beta}}(\bomega_1)
E_{X_2}\cdots
G_{\tilde{\beta}}(\bomega_{n-1})
E_{X_{n}}
\Bigg]\no
 &\times \mathcal{R}
\bigg[
E_{X_{n}}
G_{\tilde{\beta}}(\bomega_{n-1})^*
 E_{X_{n-1}}\cdots
 G_{\tilde{\beta}}(\bomega_1)^*
E_{X_1}
\bigg]\no
\unrhd& 
\sum_{X_1, X_{n}\in \wedge^{\Md}\Lambda}
\sum_{P=X_1\cdots X_{n}\in \mathscr{P}_{X_1X_{n}}^{(\Md)}[n-1]}
\int d \nu_{\bphi, \bphi'; \tilde{\beta}}^{(n-1)}\no
&\times \mathcal{L}
\bigg[
E_{X_1}
G_{\tilde{\beta}}(\bomega_1)
E_{X_2}\cdots
 G_{\tilde{\beta}}(\bomega_{n-1})
E_{X_{n}}
\bigg]\no
 &\times \mathcal{R}
\bigg[
E_{X_{n}}
G_{\tilde{\beta}}(\bomega_{n-1})^*
E_{X_{n-1}}\cdots
 G_{\tilde{\beta}}(\bomega_1)^*
E_{X_1}
\bigg]
   \label{InqCKernel}
\end{align} 
w.r.t. $\mathscr{L}^2(\Fock_{\mathrm{e}, \Md})_+$.  $\Box$
  \medskip\\

Let $\psi, \vphi\in \mathfrak{C}_M\backslash \{0\}$.
By (\ref{L2CONSEX}), we  can express these as 
\begin{align*}
\psi&=\sum_{X, Y\in
 \wedge^{\Md}\Lambda}\int^{\oplus}_{\mathcal{Q}}\psi_{XY}(\bphi)|e_X\ra \la
 e_Y|\, d \bphi,\\
\vphi&=\sum_{X, Y\in \wedge^{\Md}\Lambda}\int_{\mathcal{Q}}^{\oplus}\vphi_{X
 Y}(\bphi)|e_X\ra \la e_Y|\, d \bphi.
\end{align*} 
Since $\psi\ge 0, \vphi\ge 0$ w.r.t. $\mathfrak{C}_M$,
one obtains  $ \psi_{XX}(\bphi)=\la e_X, \psi(\bphi) e_X\ra_{\Fock_{\mathrm{e}, \Md}}\ge 0$
and  $\vphi_{XX}(\bphi)=\la e_X,
\vphi(\bphi) e_X\ra_{\Fock_{\mathrm{e}, \Md}}\ge 0$ for all  $X\in \wedge^{\Md}\Lambda$
which imply $\psi_{XX}(\bphi)\ge 0$ and $\vphi_{XX}(\bphi)\ge 0$ for all
$X\in
\wedge^{\Md} \Lambda$ and a.e. $\bphi$. In particular, since both $\psi$ and $\vphi$ are non-zero,
there exist $X, Y\in \wedge^{\Md} \Lambda$ and $\mathcal{S}_X,
\mathcal{S}_Y\subseteq \mathcal{Q}$  with non-vanishing Lebesgue measures
such  that $\psi_{XX}(\bphi)>0$ on $\mathcal{S}_X$ and
$\vphi_{YY}(\bphi)>0$ on $\mathcal{S}_Y$\footnote{
Assume that $\psi_{XX}(\cdot)=0$ for all $X\in \wedge^{\Md}\Lambda$ as a
vector in $L^2(\mathcal{Q})$. Then we have 
$
\Tr[\psi(\bphi)]=\sum_{X\in \wedge^{\Md}\Lambda} \psi_{XX}(\bphi)=0
$
, which implies that  $\psi=0$. This is a contradiction. Thus, there exists an
$X\in \wedge^{\Md} \Lambda$ such that $\psi_{XX}(\cdot)\neq 0$ as a
vector in $L^2(\mathcal{Q})$. Thus, 
there exists a measurable set $\mathcal{S}_X$ with $|\mathcal{S}_X|>0$
such that $\psi_{XX}(\bphi)>0$ for all $\bphi\in \mathcal{S}_X$.
}.
Then one obtains the following:
\begin{coro}\label{LwCNB}
It holds that 
\begin{align} 
\big\la \vphi, \mathscr{C}_{n,
 \beta}\psi\big\ra
\ge& 
\sum_{P\in \mathscr{P}_{Y X}^{(\Md)}[n-1]}
\int_{\mathcal{S}_Y\times \mathcal{S}_X} d \bphi\, d \bphi' \int  d \nu_{\bphi, \bphi';\tilde{\beta}}^{(n-1)}
\,  \vphi_{Y Y}(\bphi)\psi_{X X}(\bphi')\no
&\times \bigg|
\Big\la e_{Y},
\mathrm{\mathscr{G}}_{
\tilde{\beta}}^{(\Md)}\Big(P, \{\bomega_j\}_{j=1}^{n-1}\Big)
e_{X}\Big\ra_{\Fock_{\mathrm{e}, \Md}}
\bigg|^2.  \label{InqC2}
\end{align}
\end{coro} 
{\it Proof.} By (\ref{InqCKernel2}),  we have 
\begin{align}
&\big\la \vphi, \mathscr{C}_{n,
 \beta}\psi\big\ra\no
\ge& \sum_{X_1, X_{n}\in \wedge^{\Md}\Lambda}
\sum_{P\in \mathscr{P}_{X_1 X_{n}}^{(\Md)}[n-1]}
\int_{\mathcal{Q}\times \mathcal{Q}} d \bphi\,  d \bphi'
\int  d \nu_{\bphi, \bphi'; \tilde{\beta}}^{(n-1)}\, 
 \vphi_{X_1 X_1}(\bphi)\psi_{X_{n} X_{n}}(\bphi')\no
&\times \bigg|
\Big\la e_{X_1}, \mathrm{\mathscr{G}}_{
\tilde{\beta}}^{(\Md)}\Big(P, \{\bomega_j\}_{j=1}^{n-1}\Big)
e_{X_{n}}\Big\ra_{\Fock_{\mathrm{e}, \Md}}
\bigg|^2  \no
\ge & \mbox{ RHS of (\ref{InqC2})}.\ \ \ \Box
\end{align} 
\medskip

\begin{flushleft}
{\bf Conclusion:}
By Corollary \ref{LwCNB}, to show that $\{\mathscr{C}_{n, \beta}\}$ is ergodic, it suffices to find
some $n$ and $\beta$ such that  the RHS of (\ref{InqC2}) is strictly
positive. $\diamondsuit$
\end{flushleft} 

\medskip

 For all $\{x, y\}\in E$ and $z\in \Lambda$, set
\begin{align}
a_z=a_z(\{x, y\})=\sqrt{2}\omega_0^{-1/2} 
(g_{xz}-g_{yz}).
\end{align} 
Let 
\begin{align}
\mathcal{Y}=\bigg\{
(\bphi, \bphi')\in \mathcal{Q}\times \mathcal{Q}\, \bigg|\, \exists \{x,
 y\}\in E\ \mbox{s.t. }  \sum_{z\in
 \Lambda} a_z(\{x, y\}) (q_z-q_z') \in 2\pi \BbbZ 
\bigg\}.
\end{align}
Clearly,  $\mathcal{Y}$ is a set of Lebesgue measure $0$. 
Let 
\begin{align}
W_{\beta}=\Bigg\{
\bfphi\in A\, \Bigg|\, \max_{s\in [0, 1]} |\balpha(s)({\bfphi})| \le \beta^{-1/4}
\Bigg\}.
\end{align}
Note that $\int_{W_{\beta}} D{\boldsymbol \alpha}>0$ for sufficiently
small $\beta>0$,  since $\cup_{\beta >0} W_{\beta}=A$.
In Appendix \ref{PfC}, we will show the following:
\begin{Prop}\label{RepConnectivity2}(Connectivity)
 Let $P\in \mathscr{P}_{XY}^{(\Md)}[L]$. 
Let $(\bphi, \bphi_1), (\bphi_1, \bphi_2), \dots,\\ (\bphi_{L-1},
 \bphi')\in \mathcal{Y}^c$, the complement of $\mathcal{Y}$.
Then  there exist $\beta_*>0$ and $\Gamma_*>0$ such that  for  
all $\beta\in (0,\beta_*)$
and  $\bfphi_1, \bfphi_2\,  \dots, \bfphi_{L}\in W_{ \beta}$,
 we have
\begin{align}
\Bigg|\beta^{-L}\Big\la e_X,  \mathrm{\mathscr{G}}_{
\beta}^{(\Md)}\Big(P, \{\bomega_j(\bfphi_j)\}_{j=1}^{L}\Big)
 e_Y
\Big\ra_{\Fock_{\mathrm{e}, \Md}}\Bigg| \ge \Gamma_*.
\end{align} 
Note that $\beta_*$ and $\Gamma_*$ depend on $\bphi, \bphi_1, \dots,
 \bphi_{L-1}, \bphi'$.
\end{Prop} 
{\it Proof.} See Appendix \ref{PfC}. $\Box$

\subsection{Completion of proof of Theorem \ref{Ergodicity}}
By Propositon \ref{ConnectFermi}, we can take $n\in \BbbN$ such that 
$\mathscr{P}_{YX}^{(\Md)}[n-1]\neq \emptyset$.
Let $(\bphi, \bphi_1), \dots, (\bphi_{n-2}, \bphi')\in \mathcal{Y}^c$.
For all $P\in \mathscr{P}_{YX}^{(\Md)}[n-1],\
 \beta\in (0, \beta_*)$ and
  $\bfphi_1, \dots,  \bfphi_{n-1}\in
 W_{\beta}
$,  the term 
\begin{align*}
\bigg|
\Big\la e_{Y}, \mathrm{\mathscr{G}}_{\tilde{\beta}}
^{(\Md)}\Big(P, \{\bomega_j(\bfphi_j)\}_{j=1}^{n-1}\Big)\, 
e_{X}\Big\ra_{\Fock_{\mathrm{e}, \Md}}
\bigg|^2
\end{align*} 
 is strictly positive  by Proposition \ref{RepConnectivity2}.
Thus,  it holds that 
\begin{align}
 &\int_0^1 d\beta\int d \mu_{\bphi, \bphi_1;
 \beta}({\boldsymbol \vphi}_1)\, d \mu_{\bphi_1, \bphi_2;
 \beta}({\boldsymbol \vphi}_2)
\cdots 
d\mu_{\bphi_{n-2}, \bphi';
 \beta}({\boldsymbol \vphi}_{n-1})\no
\times& \exp\Bigg[
-\sum_{j=1}^{n-1} \int_0^{\beta} d s \mathcal{V}\big(\bomega_j(s)(\boldsymbol \vphi_j)\big)
\Bigg]
\bigg|
\Big\la e_{Y}, \mathrm{\mathscr{G}}_{\tilde{\beta}}
^{(\Md)}\Big(P, \{\bomega_j(\bfphi_j)\}_{j=1}^{n-1}\Big)\, 
e_{X}\Big\ra_{\Fock_{\mathrm{e}, \Md}}
\bigg|^2\no
&>0 \label{KBP}
\end{align} 
for all  $(\bphi, \bphi_1), \dots, (\bphi_{n-2}, \bphi')\in \mathcal{Y}^c$.
Let $\mathscr{K}_{n, \beta}$ be the RHS of (\ref{InqC2}). By (\ref{DefNot})
and (\ref{KBP}), we have 
$\int_0^1\mathscr{K}_{n, \beta} d\beta>0$. Since $\mathscr{K}_{n, \beta}$ is
continuous in $\beta$, there exists a $\beta_0>0$ such that
$\mathscr{K}_{n, \beta_0}$ is strictly positive.
Hence,  
$\{\mathscr{C}_{n, \beta}\}$ is ergodic.
   $\Box$

\subsection{Proof of Theorem  \ref{Uniqueness}}\label{DetailedGS}

By Corollary \ref{SPositive}  and Theorem \ref{Faris}, the ground state of $\mathbb{H}_M$ is unique and
strictly positive w.r.t. $\mathfrak{C}_M$.

(i) immediately follows from Theorem  \ref{PositiveGS}.

Because $H_M$ commutes with $S_{\mathrm{tot}}^2$ and because the ground
 state of $H_M$  is unique, we  obtain (ii). 

By an argument similar to  that for (\ref{InnerSpin}), we have 
\begin{align}
\big\la \psi_M, S_{x+}S_{y-}\psi_M\big\ra=\gamma_x\gamma_y \big\la
 \psi_M, \mathcal{L}(c_xc_y^*) \mathcal{R}((c_xc_y^*)^*)\psi_M\big\ra.
\end{align}
Since  $\psi_M$ is strictly positive and  $\mathcal{L}(c_xc_y^*)
\mathcal{R}((c_xc_y^*)^*) \unrhd 0$ w.r.t. $\mathfrak{C}_M$,
we conclude (iii).
 $\Box$

\section{Proof of Theorem \ref{HHmodel}} \label{InfB}

\subsection{Gaussian domination}

In this section, we assume {\bf (B. 1)}, {\bf (B. 2)} and {\bf (B. 3)}.

In the previous sections, we considered the Hamiltonian in the
$M$-subspace.
Here,  we will study the Hamiltonian  in the full space
$\mathfrak{E}
\otimes \mathfrak{P}$. In this case, we can still  define the Lang--Firsov
transformation $\mathscr{U}$ and the hole-particle  transformation
$\mathscr{W}$ as before. Let us define $\mathbb{H}$ by 
\begin{align}
\mathbb{H} =\mathscr{WU} H \mathscr{U}^* \mathscr{W}^*
+\sum_{x\in \Lambda}\mu_x n_x-\frac{1}{2}\sum_{x, y\in \Lambda} V_{xy}.
\end{align} 
We can confirm that 
\begin{align}
\mathbb{H}=-T_{+g, \uparrow}-T_{-g, \downarrow}+\tU +H_{\mathrm{p}},
\end{align} 
where $T_{\pm, \sigma}$ and $\tU$ are given in Subsections \ref{SecLF} and
 \ref{SecHP},  respectively.

For each $\mb{h}=\{h_x\}_{x\in \Lambda}\in \BbbR^{|\Lambda|}$, let 
\begin{align}
\tU(\mb{h})=\frac{1}{2}\sum_{x, y \in \Lambda}U_{\ef, xy}\big(n_{x\uparrow}-n_{x\downarrow}+h_x\big)\big(n_{y\uparrow}-n_{y\downarrow}+h_y\big).
\end{align} 
We introduce a new Hamiltonian  given by the following: 
\begin{align}
\mathbb{H}(\mb{h})=-T_{+g, \uparrow}-T_{-g,
 \downarrow}+\tU(\mb{h})+H_{\mathrm{p}}. \label{NewHami}
\end{align} 
Note that 
\begin{align}
\mathbb{H}=\mathbb{H}(\mb{0}).
\end{align} 
The main purpose in this subsection is to show the following:
\begin{Thm}\label{Gaussian}
Let 
$
\mathcal{Z}_{\beta, \vepsilon}(\mb{h})=\Tr\Big[
\ex^{-\beta \mathbb{H}(\mb{h})}
\ex^{-\vepsilon H_{\mathrm{p}}}
\Big].
$
We have  $\mathcal{Z}_{\beta, \vepsilon}(\mb{h}) \le \mathcal{Z}_{\beta, \vepsilon}(\mb{0})$ for all $\mb{h}\in
 \BbbR^{|\Lambda|}$ and $\vepsilon>0$.
\end{Thm} 
\begin{rem}
{\rm
We introduced $\ex^{-\vepsilon H_{\mathrm{p}}}$ in
 $\mathcal{Z}_{\beta, \vepsilon}(\mathbf{h})$ for the  following reason:
the factor $\ex^{-\vepsilon H_{\mathrm{p}}}$ enables us to interchange a
 limit operation and a trace operation in the final step of the proof.
$\diamondsuit$
}
\end{rem} 
\subsubsection{Auxiliary lemmas}

Let $T=T_{+g, \uparrow}+T_{-g, \downarrow}$. Under the
identification 
\begin{align}
\mathfrak{E}\otimes \mathfrak{P}=\int^{\oplus}_{\mathcal{Q}}
 \Fock_{\mathrm{e}}\otimes \Fock_{\mathrm{e}} d\bphi,
\end{align} 
 we have 
\begin{align}
T=\int_{\mathcal{Q}}^{\oplus} T(\bphi) d\bphi,\ \ \ \ 
 T(\bphi)={\bf T}_{+g}(\bphi)\otimes \one+\one \otimes {\bf T}_{-g}(\bphi),
\end{align} 
where ${\mb T}_{\pm g}(\bphi)$ is defined by (\ref{Tg}).

\begin{lemm}\label{TraceLemma}
Let $K=-T+H_{\mathrm{p}}$.
Let
\begin{align}
\mathcal{Z}_{\beta, n, \vepsilon}(\mb{h})=\Tr\bigg[
\Big(
\ex^{-\beta K /n}\ex^{-\beta \tU(\mb{h})/n}
\Big)^n \ex^{-\vepsilon H_{\mathrm{p}}}
\bigg],\ \ n\in \BbbN,\ \ \vepsilon>0.
\end{align}
Let us introduce the following  notation:
\begin{align}
&\int d\nu^{(n+1)}_{\bphi, \bphi'; \beta, \vepsilon}F(\bomega_1,\dots,
 \bomega_{n+1})\no
:=&\int_{\mathcal{Q}^{n}}\prod_{j=1}^n d\bphi_j \int d\mu_{\bphi,
 \bphi_1; \beta}\int d \mu_{\bphi_1, \bphi_2; \beta}\cdots
\int d \mu_{\bphi_{n-1}, \bphi_n; \beta}
 \int
 d\mu_{\bphi_{n}, \bphi'; \vepsilon}\no
&\times \exp\Bigg\{-\sum_{j=1}^{n}\int_0^{\beta}ds
\mathcal{V}(\bomega_j(s))
-\int_0^{\vepsilon}ds \mathcal{V}(\bomega_{n+1}(s))
\Bigg\}
F\big(\bomega_1,\dots, \bomega_{n+1}\big).
\end{align} 
Then, setting $\tilde{\beta}=\beta/n$,  we have  
\begin{align}
&\mathcal{Z}_{\beta, n, \vepsilon}(\mb{h})\no
=&(4\pi)^{-n|\Lambda|/2}
\int_{\BbbR^{n |\Lambda|}}\prod_{j=1}^n  d\mb{k}_j
\int_{\mathcal{Q}}d\bphi
\int d\nu_{\bphi, \bphi; \beta, \vepsilon}^{(n+1)}
\ex^{-\im \sum_{j=1}^n  \mb{k}_j\cdot \mb{h}}\ex^{-\sum_{j=1}^n
 \mb{k}^2_j/4}\no
&\times \Tr_{\Fock_{\mathrm{e}}\otimes \Fock_{\mathrm{e}}}
\Bigg[
\prod_{j=1}^{{n}\atop{\longrightarrow}} \Bigg(
\prod_{0}^{{\tilde{\beta}}\atop{\longrightarrow}}
\ex^{T(\bomega_j(s))ds}
\ex^{\im \sum_{x, y\in \Lambda}
\tilde{\beta}
k_{j x}U_{\ef, xy}(\mathsf{n}_{y}\otimes \one - \one \otimes \mathsf{n}_{y})
}
\Bigg)
\Bigg].
\end{align} 
\end{lemm} 
{\it Proof.} By the Trotter--Kato product formula, we have
\begin{align}
\ex^{-\beta K }(\bphi, \bphi')=\int d\mu_{\bphi, \bphi';
 \beta}
\Bigg(
\prod_0^{{\beta}\atop{\longrightarrow}} \ex^{T(\bomega(s))ds} 
\Bigg) \ex^{-\int_0^{\beta}ds \mathcal{V}(\bomega(s))}.\label{Kernel}
\end{align} 
Let 
\begin{align}
\mathscr{I}_{n, \beta, \vepsilon}=\Big(
\ex^{-\beta K /n}\ex^{-\beta \tU(\mb{h})/n}
\Big)^n
\ex^{-\vepsilon H_{\mathrm{p}}}.
\end{align} 
By (\ref{Kernel}), the kernel operator  of $\mathscr{I}_{n, \beta,
\vepsilon}$ is obtained by the following observation:
\begin{align}
&\mathscr{I}_{n, \beta, \vepsilon}(\bphi_0, \bphi_{n+1})\no
=&\int_{\mathcal{Q}^n}\prod_{j=1}^n d\bphi_j
\Bigg(
\prod_{j=1}^{{n}\atop{\longrightarrow}} \ex^{-\beta K /n}(\bphi_{j-1}, \bphi_j)
\ex^{-\beta \tU(\mb{h})/n}
\Bigg)\ex^{-\vepsilon H_{\mathrm{p}}}(\bphi_n, \bphi_{n+1})\no
=&\int_{\mathcal{Q}^n} \prod_{j=1}^n d\bphi_j
\int d\mu_{\bphi_0, \bphi_1; \tilde{\beta}} \cdots \int d\mu_{\bphi_n,
 \bphi_{n+1}; \vepsilon}
\ex^{
-\sum_{j=1}^n\int_0^{\tilde{\beta}} ds
 \mathcal{V}(\bomega_j(s))-\int_0^{\vepsilon}
ds \mathcal{V}(\bomega_{n+1}(s))
}\no
&\times \prod_{j=1}^{{n}\atop{\longrightarrow}}
\Bigg\{
\Bigg(
\prod_0^{{\tilde{\beta}}\atop{\longrightarrow}} \ex^{T(\bomega_j(s))ds}
\Bigg)
\ex^{-\tilde{\beta} \tU(\mb{h})}
\Bigg\}\no
=&\int d\nu_{\bphi_0, \bphi_{n+1}; \beta, \vepsilon}^{(n+1)}
 \prod_{j=1}^{{n}\atop{\longrightarrow}}
\Bigg\{
\Bigg(
\prod_0^{{\tilde{\beta}}\atop{\longrightarrow}} \ex^{T(\bomega_j(s))ds}
\Bigg)
\ex^{-\tilde{\beta} \tU(\mb{h})}
\Bigg\}.
\end{align} 
Thus,  we have 
\begin{align}
\mathcal{Z}_{\beta, n, \vepsilon}(\mb{h})&=\Tr[\mathscr{I}_{n, \beta,
 \vepsilon}]=\int_{\mathcal{Q}}d\bphi \Tr_{\Fock_{\mathrm{e}} \otimes \Fock_{\mathrm{e} }}\Big[
\mathscr{I}_{n, \beta, \vepsilon}(\bphi, \bphi)
\Big]\no
&=\int_{\mathcal{Q}}d\bphi \int d\nu_{\bphi, \bphi; \beta,
 \vepsilon}^{(n+1)}
\Tr_{\Fock_{\mathrm{e}}\otimes \Fock_{\mathrm{e}}}\Bigg[
\prod_{j=1}^{{n}\atop{\longrightarrow}}
\Bigg\{
\Bigg(
\prod_0^{{\tilde{\beta}}\atop{\longrightarrow}} \ex^{T(\bomega_j(s))ds}
\Bigg)
\ex^{-\tilde{\beta} \tU(\mb{h})}
\Bigg\}
\Bigg].
\end{align} 
Finally,    applying the following identity
\begin{align}
\ex^{-\tilde{\beta} \tU(\mb{h})}
=(4\pi)^{-|\Lambda|/2}\int_{\BbbR^{|\Lambda|}} d\mb{k}
\ex^{-\im \mb{h}\cdot \mb{k}}\ex^{-\mb{k}^2/4}
\ex^{\im \sum_{x, y\in \Lambda}\tilde{\beta}U_{\ef, xy}k_x (n_{y\uparrow}-n_{y\downarrow})},
\end{align} 
we obtain the assertion in the lemma. $\Box$

\begin{lemm}
We have 
\begin{align}
&\mathcal{Z}_{\beta, n, \vepsilon}(\mb{h})\no
=&(4\pi)^{-n|\Lambda|/2}
\int_{\BbbR^{n |\Lambda|}}\prod_{j=1}^n  d\mb{k}_j
\int_{\mathcal{Q}}d\bphi
\int d\nu^{(n+1)}_{\bphi, \bphi; \beta, \vepsilon} 
\ex^{-\im \sum_{j=1}^n  \mb{k}_j\cdot \mb{h}}\ex^{-\sum_{j=1}^n  \mb{k}^2_j/4}
\no
&\times
\Bigg|
\Tr_{\Fock_{\mathrm{e}}}\Bigg[
\prod_{j=1}^{{n}\atop{\longrightarrow}}
\Bigg(
\prod_{0}^{{\tilde{\beta}}\atop{\longrightarrow}}
\ex^{T_{+g}(\bomega_j(s))ds}
\ex^{\im  \sum_{x, y\in \Lambda}\tilde{\beta}k_{j x}U_{\ef, xy} \mathsf{ n}_y}
\Bigg)
\Bigg]
\Bigg|^2. \label{TraceF2}
\end{align} 
\end{lemm} 
{\it Proof.}
Note that $\Tr[A\otimes B]=\Tr[A] \Tr[B]$.
By Lemma \ref{TraceLemma}, we immediately have
\begin{align}
&\mathcal{Z}_{\beta, n, \vepsilon}(\mb{h})\no
=&(4\pi)^{-n |\Lambda|/2}
\int_{\BbbR^{n |\Lambda|}}\prod_{j=1}^n d\mb{k}_j
\int_{\mathcal{Q}} d\bphi
\int d\nu^{(n+1)}_{\bphi, \bphi; \beta, \vepsilon}
\ex^{-\im \sum_{j=1}^n \mb{k}_j\cdot \mb{h}}\ex^{-\sum_{j=1}^n  \mb{k}^2_j/4}
\no
&\times
\Tr_{\Fock_{\mathrm{e}}}\Bigg[
\prod_{j=1}^{{n}\atop{\longrightarrow}}
\Bigg(
\prod_{0}^{{\tilde{\beta}}\atop{\longrightarrow}}
\ex^{T_{+g}(\bomega_j(s))ds}
\ex^{+\im  \sum_{x, y\in \Lambda} \tilde{\beta}k_{j x}U_{\ef, xy} \mathsf{n}_y}
\Bigg)
\Bigg]\no
&\times 
\Tr_{\Fock_{\mathrm{e}}}\Bigg[
\prod_{j=1}^{{n}\atop{\longrightarrow}}
\Bigg(
\prod_{0}^{{\tilde{\beta}}\atop{\longrightarrow}}
\ex^{T_{-g}(\bomega_j(s))ds}
\ex^{-\im  \sum_{x, y\in \Lambda}\tilde{\beta}k_{j x}U_{\ef, xy} \mathsf{n}_y}
\Bigg)
\Bigg]
.
\end{align} 

Let $\Theta$ be a conjugation in $\Fock_{\mathrm{e}}$ defined by $\Theta c_{x_1}^* \cdots
c_{x_{N}}^*\Omega=c_{x_1}^*\cdots c_{x_{N}}^* \Omega$,
 where $\Omega$ is the Fock vacuum in $\Fock_{\mathrm{e}}$.
Noting that $\Theta c_x \Theta =c_x$, we have 
$
\Theta T_{ -g}(\bomega(s))\Theta=T_{+g}(\bomega(s))$ and
$ \Theta \mathsf{n}_x\Theta =\mathsf{n}_x.
$
Thus,  it holds that 
\begin{align}
\Theta \prod_{0}^{{\tilde{\beta}}\atop{\longrightarrow}} \ex^{T_{-g}(\bomega(s))ds}\Theta
&=\prod_{0}^{{\tilde{\beta}}\atop{\longrightarrow}} \ex^{T_{ +g}(\bomega(s))ds},\\
\Theta \ex^{-\im  \sum_{x, y\in \Lambda}
 \tilde{\beta}k_{j x} U_{\ef, xy}\mathsf{n}_y }
\Theta
&=\ex^{+\im  \sum_{x, y\in \Lambda}
 \tilde{\beta}k_{j x} U_{\ef, xy}\mathsf{n}_y }.
\end{align} 
Hence,  using the fact that $\Tr[A]=(\Tr[\Theta A \Theta])^*$, we  observe
that 
\begin{align}
&\Tr_{\Fock_{\mathrm{e}}}\Bigg[
\prod_{j=1}^{{n}\atop{\longrightarrow}}
\Bigg(
\prod_{0}^{{\tilde{\beta}}\atop{\longrightarrow}} \ex^{T_{ -g}(\bomega(s))ds}
\ex^{-\im \sum_{x, y\in \Lambda} \tilde{\beta}k_{j x}U_{\ef, xy}\mathsf{n}_y}
\Bigg)
\Bigg]\no
=&
\Bigg\{
\Tr_{\Fock_{\mathrm{e}}}\Bigg[\Theta
\prod_{j=1}^{{n}\atop{\longrightarrow}}
\Bigg(
\prod_{0}^{{\tilde{\beta}}\atop{\longrightarrow}} \ex^{T_{ -g}(\bomega(s))ds}
\ex^{-\im \sum_{x, y\in \Lambda} \tilde{\beta}k_{j x}U_{\ef, xy}\mathsf{n}_y}
\Bigg)
\Theta
\Bigg]
\Bigg\}^*\no
=&
\Bigg\{
\Tr_{\Fock_{\mathrm{e}}}\Bigg[
\prod_{j=1}^{{n}\atop{\longrightarrow}}
\Bigg(
\prod_{0}^{{\tilde{\beta}}\atop{\longrightarrow}} \ex^{T_{ +g}(\bomega(s))ds}
\ex^{+\im \sum_{x, y\in \Lambda} \tilde{\beta}k_{j x}U_{\ef, xy}\mathsf{n}_y}
\Bigg)
\Bigg]
\Bigg\}^*.
\end{align} 
This completes the proof. $\Box$

\subsubsection{Proof of Theorem \ref{Gaussian}}

Remark that except for   $\ex^{-\im \sum_{j=1}^n\mb{k}_j\cdot \mb{h}}$,
all  factors of
the integrand in (\ref{TraceF2}) are positive. Thus,  
$|\mathcal{Z}_{\beta, n, \vepsilon}(\mb{h})| \le
\mathcal{Z}_{\beta, n, \vepsilon}(\mb{0})$. As  $n\to \infty$,
$ \mathcal{Z}_{\beta, n, \vepsilon}(\mathbf{h})$ converges to
$ \mathcal{Z}_{\beta, \vepsilon}(\mathbf{h})$  by Lemma
\ref{Grumm} below.
Thus,  we have $\mathcal{Z}_{\beta,  \vepsilon}(\mb{h}) \le
\mathcal{Z}_{\beta, \vepsilon}(\mb{0})$.
 $\Box$

\begin{lemm}\label{Grumm}
We denote by  $\mathscr{L}^1(\mathfrak{X})$  the ideal of all trace class
 operators on  a Hilbert space $\mathfrak{X}$.
Let $A_n, A \in \mathscr{B}(\mathfrak{X})$ and $B_n, B
 \in \mathscr{L}^1(\mathfrak{X})$ such that $A_n$ converges to $A$ strongly and 
$\|B_n-B\|_1\to 0$ as $n\to \infty$, where $\|\cdot\|_1$ is the trace
 norm. Then $\|A_n B_n-AB\|_1\to 0$ as $n\to \infty$. 
\end{lemm} 
{\it Proof.} See \cite[Chap. 2, Example 3]{Simon2}. $\Box$

\subsection{
Completion of   proof of Theorem \ref{HHmodel}
 }

 We define the Duhamel two-point function as 
\begin{align}
(\!(A, B)\!)_{\beta, \Lambda}=\mathcal{Z}_{\beta}^{-1}\int_0^1 dx \Tr
\Big[
\ex^{-x \beta\mathbb{H}}A \ex^{-(1-x)\beta\mathbb{H}}
B
\Big].
\end{align} 

\begin{Thm} \label{SusInq}
Let $\sigma_x=n_{x\uparrow}-n_{x\downarrow}$.
For all $\mb{h}\in \Bbb{C}^{|\Lambda|}$,  we have  
\begin{align}
\Big(\!\!\Big(
\big\la \Sig , \mb{\Ue} \mb{h}\big\ra^*,  \big\la \Sig , \mb{\Ue} \mb{h}\big\ra
\Big)\!\!\Big)_{\beta, \Lambda} \le \beta^{-1} \big\la \mb{h},
 \mb{\Ue}\mb{h}\big\ra,
\label{Susp}
\end{align} 
where $\la \Sig, \mb{\Ue}\mb{h}\ra=\sum_{x,y\in
 \Lambda} U_{\ef, xy}\sigma_x h_y$.
\end{Thm} 
{\it Proof.} Let $\lambda\in \BbbR$. We note
\begin{align}
\mathbb{H}(\lambda \mb{h})&=\mathbb{H}+\delta \tU(\lambda
 \mb{h}),\\
\delta\tU(\lambda \mb{h})&=\tU(\lambda
 \mb{h})-\tU(\mb{0})= \lambda \la \Sig , \mb{\Ue}
 \mb{h}\ra+\frac{\lambda^2}{2} \la \mb{h}, \mb{\Ue}\mb{h}\ra.
\end{align} 
By the Duhamel formula, we have the norm-convergent expansion:
\begin{align}
\ex^{-\beta \mathbb{H}(\lambda\mb{h})}&=\sum_{n=0}^{\infty}
 \mathcal{D}_n(\lambda),\\
\mathcal{D}_n(\lambda)&=(-\beta)^{n} \int_{S_n(1)}
\ex^{-s_1 \beta \mathbb{H}} \delta \tU(\lambda\mb{h}) \cdots \ex^{-s_n \beta \mathbb{H}} \delta
 \tU(\lambda \mb{h})
\ex^{-(1-\sum_{j=1}^n s_j)\beta \mathbb{H}}.
\end{align} 
By Lemma \ref{Grumm}, we have 
\begin{align}
\mathcal{Z}_{\beta, \vepsilon}(\lambda \mb{h})=\sum_{n=0}^{\infty}\Tr
\Big[
\mathcal{D}_n(\lambda)\ex^{-\vepsilon H_{\mathrm{p}}}
\Big].
\end{align} 
  Note that 
\begin{align}
\Tr\Big[\mathcal{D}_1(\lambda)\ex^{-\vepsilon
 H_{\mathrm{p}}}\Big]=\frac{\lambda^2}{2}\la \mb{h},
 \mb{\Ue}\mb{h}\ra\Tr\Big[
\ex^{-\beta \mathbb{H} }\ex^{-\vepsilon H_{\mathrm{p}}}
\Big]
\end{align} 
and,  by Theorem \ref{Gaussian},
\begin{align}
\frac{\mathcal{Z}_{\beta, \vepsilon}(\mb{0})-\mathcal{Z}_{\beta, \vepsilon}(\lambda
 \mb{h})}{\lambda^2}\ge 0.
\end{align} 
Hence,  letting $\lambda\to 0$, it follows s that 
\begin{align}
 &\frac{\beta}{2}\la \mb{h}, \mb{\Ue}\mb{h\ra}\Tr\Big[
\ex^{-\beta \mathbb{H} }\ex^{-\vepsilon H_{\mathrm{p}}}
\Big]\no
&-\beta^2 \int_0^1ds_1 \int_0^{1-s_1}
ds_2
\Tr\Big[
\ex^{-s_1 \beta \mathbb{H}}\la  \Sig,
 \mb{\Ue}\mb{h}\ra\ex^{-s_2 \beta \mathbb{H}}
\la \Sig, \mb{\Ue}\mb{h}\ra\no
&\ \ \ \ \times \ex^{-(1-s_1-s_2)\beta \mathbb{H}}
\ex^{-\vepsilon H_{\mathrm{p}}}
\Big]\ge 0.\label{RareInq}
\end{align} 
By applying Lemma \ref{Grumm} again, we have $\lim_{\vepsilon\to +0}
\Tr[ \ex^{-\beta \mathbb{H}} \ex^{-\vepsilon H_{\mathrm{p}}}]=\mathcal{Z}_{\beta}
$ and 
\begin{align}
&\mbox{the second term in (\ref{RareInq}) }
\no
\to & \frac{\beta^2}{2}\int_0^1dx \Tr\Big[
\la \Sig, \mb{\Ue} \mb{h}\ra \ex^{-x \beta \mathbb{H}}
\la \Sig, \mb{\Ue}\mb{h}\ra \ex^{-(1-x)\beta \mathbb{H}}
\Big]  \ \mbox{as }\ \  \vepsilon \to +0.
\end{align} 
Thus,  we obtain (\ref{Susp})  for $\mb{h}$ real-valued.
To extend this to complex-valued $\mb{h}$'s, we just note that,
if $A=A_{\mathrm{R}}+\im A_{\mathrm{I}}$ with $A_{\mathrm{R}}^* =
A_{\mathrm{R}},\ A_{\mathrm{I}}^* =A_{\mathrm{I}}$, we have $ 
(\!(A^*, A)\!)_{\beta, \Lambda}=(\!(A_{\mathrm{R}},
A_{\mathrm{R}})\!)_{\beta, \Lambda}+(\!(A_{\mathrm{I}}, A_{\mathrm{I}})\!)_{\beta, \Lambda}$. 
$\Box$
\medskip\\

To finish proof of  Theorem \ref{HHmodel}, we note that 
\begin{align}
\Big(
\la \delta {\bf n}, \mb{U}_{\mathrm{eff}} \mb{h}\ra^*, \la \delta
 \mb{n}, \mb{U}_{\mathrm{eff}} \mb{h}\ra
\Big)_{\beta, \Lambda}
=
\Big(\!\!\Big(
\big\la \Sig , \mb{\Ue} \mb{h}\big\ra^*,  \big\la \Sig, \mb{\Ue} \mb{h}\big\ra
\Big)\!\!\Big)_{\beta, \Lambda}.
\end{align} 
Thus, by the Fourier transformation, we obtain Theorem \ref{HHmodel}. $\Box$

\appendix

\section{Operator inequalities associated with the Hilbert cone}\label{GeneralSC}

 Let $\mathfrak{X}$ be a complex Hilbert space and $\mathfrak{X}_+$ be a
 Hilbert cone in $\mathfrak{X}$.

\begin{Prop}\label{Mono}
Let $A, B$ be self-adjoint positive 
 operators on $\mathfrak{X}$. Suppose 
 that 
\begin{itemize}
\item[{\rm (i)}] $\ex^{-\beta A}\unrhd 0$ w.r.t. $\mathfrak{X}_+$ for
	     all $\beta\ge 0$;
\item[{\rm (ii)}] $A \unrhd B$ w.r.t. $\mathfrak{X}_+$;
\item[{\rm (iii)}] $C=A-B$ is bounded.
\end{itemize} 
Then we have $\ex^{-\beta B }\unrhd \ex^{-\beta A}$
 w.r.t. $\mathfrak{X}_+$ for all $\beta\ge  0$. 
\end{Prop} 
{\it Proof.} By (ii), $C\unrhd 0$ w.r.t. $\mathfrak{X}_+$ and
$B=A-C$. By the Duhamel formula, we have the following norm-convergent expansion:
\begin{align}
\ex^{-\beta B}&=\sum_{n=0}^{\infty}D_n(\beta), \label{Duha}\\
D_n(\beta)&=\int_{S_n(\beta)} \ex^{-s_1 A}C \ex^{-s_2 A}C\cdots
 \ex^{-s_n A} C \ex^{-(\beta-\sum_{j=1}^ns_j)A},
\end{align} 
where $\int_{S_n(\beta)}=\int_0^{\beta}ds_1\int_0^{\beta-s_1}ds_2\cdots
\int_0^{\beta-\sum_{j=1}^{n-1}s_j} ds_n$ and
$D_0(\beta)=\ex^{-\beta A}$. Since $C \unrhd 0$ and $\ex^{-t
A}\unrhd 0$ w.r.t. $\mathfrak{X}_+$, it holds that
$D_n(\beta)\unrhd 0$ w.r.t. $\mathfrak{X}_+$ for all $n\ge 0$.
 Thus,  by (\ref{Duha}), we have $\ex^{-\beta B}\unrhd
 D_0(\beta)=\ex^{-\beta A}$ w.r.t. $\mathfrak{X}_+$ for all
 $\beta \ge 0$. $\Box$
\medskip

The following theorem  plays  an important role:
\begin{Thm}\label{Faris}{\rm (Perron--Frobenius--Faris)}
Let $A$ be a positive self-adjoint operator on $\mathfrak{X}$. Suppose that 
 $0\unlhd \ex^{-tA}$ w.r.t. $\mathfrak{X}_+$ for all $t\ge 0$ and $\inf
 \mathrm{spec}(A)$ is an eigenvalue.
Let $P_A$ be the orthogonal projection onto the closed subspace spanned
 by  eigenvectors associated with   $\inf
 \mathrm{spec}(A)$.
 Then the following
 are equivalent:
\begin{itemize}
\item[{\rm (i)}] 
$\dim \ran P_A=1$ and $P_A\rhd 0$ w.r.t. $\mathfrak{X}_+$.
\item[{\rm (ii)}] $0\lhd \ex^{-tA}$ w.r.t. $\mathfrak{X}_+$ for all
	     $t>0$.
\item[{\rm (iii)}] For each $x, y\in \mathfrak{X}_+\backslash\{0\}$,
there exists a $t>0$ such that $\la x, \ex^{-tA} y\ra>0$.
\end{itemize}
\end{Thm} 
{\it Proof.} See  \cite{Faris, Miyao1, ReSi4}. $\Box$

\section{Strong product integration}\label{SPI}
Let $\mathbb{C}_{n\times n}$ be the space of $n\times n$ matrices with
complex entries. Let $A(\cdot): [0, a]\to \mathbb{C}_{n\times n}$ be
continuous. Let $P=\{s_0, s_1, \dots, s_n\}$ be  a partition of $[0, a]$
and $\mu(P)=\max_j\{s_j-s_{j-1}\}$.  The {\it strong product integration of
$A$}
is defined by 
\begin{align}
\prod_0^{{a}\atop{\longrightarrow}} \ex^{A(s)ds}:=\lim_{\mu(P)\to
 0}\ex^{A(s_1)(s_1-s_0)} \ex^{A(s_2)(s_2-s_1)} \cdots   \ex^{A(s_n)(s_n-s_{n-1})}
 .
\end{align} 
Note that the limit is independent of any partition $P$.
\begin{Thm}\label{ProdInq}
It holds that 
\begin{align}
\Bigg\|
\prod_0^{{a}\atop{\longrightarrow}} \ex^{A(s)ds}-\one -\int_0^a ds A(s)
\Bigg\|\le 
\ex^{\int_0^a ds \|A(s)\|} -1 -\int_0^a ds \|A(s)\|.
\end{align} 
\end{Thm} 
{\it Proof.} See \cite{Dollard}. $\Box$

\section{Proof of Proposition \ref{RepConnectivity2}}\label{PfC}
To show Proposition \ref{RepConnectivity2}, we need two technical
lemmas.

Recall the definition of $\Phi_{\{x, y\}}(\cdot)$ given by (\ref{Phase}).

\begin{lemm} \label{PhaseLower}
Let $(\bphi, \bphi')\in \mathcal{Y}^c$, the complement
 of $\mathcal{Y}$.
  There exist $\beta_0>0$ and $C>0$ such that, for all
 $\beta \in (0, \beta_0)$ and $\bfphi\in W_{ \beta}$, 
\begin{align}
\Bigg|
\beta^{-1} \int_0^{\beta} ds \exp\Big\{
\im \Phi_{\{x, y\}}\big(\bomega(s)(\bfphi)\big)
\Big\}
\Bigg|
\ge \gamma_{xy} -C \beta^{1/4},
\end{align} 
where 
\begin{align}
\gamma_{xy}=2\bigg|
\frac{\sin\theta_{xy}}{\theta_{xy}}
\bigg|,\ \ \
\theta_{xy}=\frac{1}{2}\sum_{z\in \Lambda} a_z(\{x, y\}) (q_z'-q_z).
\end{align} 
Note that $\gamma_{xy}>0$ for all $(\bphi, \bphi')\in \mathcal{Y}^c$ and
 $\beta_0$ depends on $(\bphi, \bphi')$.
\end{lemm} 
{\it Proof.} 
Let 
\begin{align}
K_{xy}=\frac{1}{2\theta_{xy} } \ex^{\im \sum_{z\in \Lambda} a_z q_z}
\Big(
\ex^{2 \im \theta_{xy} }-1
\Big).
\end{align} 
Note that $|K_{xy}|=\gamma_{xy}$ and 
\begin{align}
K_{xy}=\beta^{-1} \int_0^{\beta} ds \exp\bigg\{
\im \Phi_{\{x, y\}}\Big(
(1-\beta^{-1}s)q_z+\beta^{-1} s q_z'
\Big)
\bigg\}.
\end{align} 
Thus, since $|\ex^{\im  a}-1|\le |a|$, we have 
\begin{align}
&\Bigg|
\beta^{-1} \int_0^{\beta} ds \exp\Big\{
\im \Phi_{\{x, y\}}\big(\bomega(s)(\bfphi)\big)
\Big\}-K_{xy}
\Bigg|\no
= &
\Bigg|
\beta^{-1} \int_0^{\beta}ds
 \exp\bigg\{
\im \Phi_{\{x, y\}}
\Big(
(1-\beta^{-1}s)q_z+\beta^{-1} s q_z'
\Big)
\bigg\}\no
&\times
\Bigg(
\exp\Big\{
\im \sqrt{\beta} \Phi_{\{x, y\}} \Big(
\balpha(s)(\bfphi)
\Big)
\Big\}-1
\Bigg)
\Bigg|\no
\le &
 \max_{s\in [0, \beta]}
\bigg|
\sqrt{\beta} \Phi_{\{x, y\}} \Big(
\balpha(s)(\bfphi)
\Big)
\bigg|\no
\le&  \beta^{1/4} \sum_{z\in \Lambda} \big|a_z(\{x, y\})\big|.
\end{align} 
This completes the proof. $\Box$

\begin{lemm}\label{Tnonvanish}
Let $(\bphi, \bphi')\in \mathcal{Y}^c$. Let $\{X, Y\} \in \wedge^{\Md}
 E$.
There exist
 $\beta_0>0$
 and $\gamma>0$ such that, for all $\beta\in (0, \beta_0)$ and $\bfphi \in W_{ \beta}$, we have 
\begin{align}
\bigg|
\bigg\la e_X, \beta^{-1}\int_0^{\beta} ds 
\mathbb{T}_{+g}\big(
\omega(s)(\bfphi)
\big) 
e_Y\bigg\ra
\bigg|
\ge \gamma.
\end{align} 
Note that $\beta_0$ and $\gamma=\gamma(\bphi, \bphi')$ depend on $(\bphi, \bphi')$.
\end{lemm} 
{\it Proof.}
 Using  standard notation of the second
quantization\footnote{
 Let $A$ be a bounded self-adjoint
operator on $\ell^{2}(\Lambda)$.  The second quantization of $A$ is
 defined by 
\begin{align}
\dG(A)_N&=\sum_{j=1}^N \one \otimes \cdots \otimes \underbrace{A}_{j \mathrm{th}} \otimes \cdots
 \otimes \one.
\end{align} 
$\dG(A)_N$ acts in $\wedge^N \ell^2(\Lambda)$.
Set $a_{xy}=\la e_x, A e_y\ra$. Then $\dG(A)_N$ can be expressed as 
\begin{align}
\dG(A)_N=\sum_{x, y\in \Lambda} a_{xy}c_x^* c_y.
\end{align} 
},
 we can write 
\begin{align}
\mathbb{T}_{+g}(\bphi)&=d\Gamma(\mathcal{T}_{+g}(\bphi))_{\Md},\label{Hoppq}\\
\mathcal{T}_{+g}(\bphi)&=\sum_{\{x,y\}\in E}  t_{xy} \exp\big\{\im \Phi_{\{x, y\}}(\bphi)\big\}
 |e_x\ra\la e_y |
\end{align} 
for all $\bphi\in \mathcal{Q}$.

Write $X, Y$ as   $X=(x_1, \dots, x_{\Md})$ and $ Y=(y_1, \dots,
y_{\Md})$.   Then,  there
exists a unique $j$ such that $\{x_j, y_j\}\in E$ and $x_i=y_i$ holds
for all $i\neq j$. By (\ref{Hoppq}), it indicates the following: 
\begin{align}
\Big\la e_X, \int_0^{\beta}d
	     s\, \mathbb{T}_{+g}\big(\bomega(s)(\bfphi)\big) e_Y\Big\ra
=&\Big\la e_{x_j}, \int_0^{\beta} ds \mathcal{T}_{+g}\big(\bomega(s)(\bfphi)\big)e_{y_j}
 \Big\ra\no
=& \int_0^{\beta} d s\, t_{x_jy_j}\,  \exp\Big\{\im
 \Phi_{\{x_j, y_j\}}\big(\bomega(s)(\bfphi)\big)
\Big\}. 
\label{Nonvanishing}
\end{align}
By  (\ref{Nonvanishing}) and Lemma
\ref{PhaseLower}, we have 
\begin{align}
&\bigg|
\bigg\la e_X, \beta^{-1}\int_0^{\beta} ds \mathbb{T}_{+g}\big(
\bomega(s)(\bfphi)
\big)
e_Y\bigg\ra
\bigg|\no
=&|t_{x_jy_j}| 
\Bigg|
\beta^{-1}\int_0^{\beta} ds \exp\bigg\{
\im \Phi_{\{x_j, y_j\}} \big(
\bomega(s)(\bfphi)
\big)
\bigg\}
\Bigg|\no
\ge & |t_{x_jy_j}|(\gamma_{x_jy_j}-C\beta^{1/4}).
\end{align} 
Thus,  we have the desired assertion. $\Box$
\medskip\\
\begin{flushleft}
{\it Completion of proof of Proposition \ref{RepConnectivity2}}
\end{flushleft} 
For each $P=X_1 X_2\cdots X_{L+1}\in \mathscr{P}_{XY}^{\Md}[L]$,
let
\begin{align}
\tau^{(\Md)}_{\beta}\Big(P, \{\bomega_j(\bfphi_j)\}_{j=1}^{L}\Big)
= \prod_{j=1}^{{L}\atop{\longrightarrow}} E_{X_j}
\int_0^{\beta}d s_j\, \mathbb{T}_{+g}\big(\bomega_j(s_j)(\bfphi_j)\big) E_{X_{j+1}}. 
\end{align}
We claim that 
\begin{align}
\mathrm{\mathscr{G}}_{
\beta}^{(\Md)}\Big(P, \{\bomega_j(\bfphi_j)\}_{j=1}^{L}\Big)
=
\tau^{(\Md)}_{\beta}\Big(P,
 \{\bomega_j(\bfphi_j)\}_{j=1}^{L}\Big)+\mathcal{O}(\beta^{L+1}). \label{GTau}
\end{align} 
Here, the error term $\mathcal{O}(\beta^{L+1})$ satisfies $\| \mathcal{O}(\beta^{L+1})\| \le C
  \beta^{L+1}$,
where $C$ is independent of $\bfphi_j$.
To see this, we observe that, by Theorem  \ref{ProdInq},  
\begin{align}
&\Bigg\|
E_{X_j}\Bigg[ G_{\beta}\big(\bomega_j(s_j)(\bfphi_j)\big)
-\int_0^{\beta}d s\,  
\mathbb{T}_{+g}\big(\bomega_j(s)(\bfphi_j)\big)
\Bigg]E_{X_{j+1}}
\Bigg\|\no
=&\Bigg\|
E_{X_j}\Bigg[ G_{\beta}\big(\bomega_j(s_j)(\bfphi_j)\big)-\one 
-\int_0^{\beta}d s\,  
\mathbb{T}_{+g}\big(\bomega_j(s)(\bfphi_j)\big)
\Bigg]E_{X_{j+1}}
\Bigg\|\no
&\le
  \bigg(
\int_0^{\beta} d s\, \big\|\mathbb{T}_{+g}
\big(\bomega_j(s)(\bfphi_j)\big)
\big\|
\bigg)^2\no
&\le \beta^2 C (\Md)^2  (\max_{x, y} |t_{xy}|)^2.
\end{align} 
 Here,  we have used the fact that
$E_{X_j}E_{X_{j+1}}=0$.

Denote $X_0=X$ and $X_{L+1}=Y$. 
By Lemma \ref{Tnonvanish}, we have 
\begin{align}
&\Bigg|\beta^{-L} \Big\la e_X, \tau_{\beta}^{\Md}\Big(P,
 \{\bomega_j(\bfphi_j)\}_{j=1}^L\Big)e_Y \Big\ra
\Bigg|\no
=& \prod_{j=1}^{L+1}
\Bigg|
\Bigg\la e_{X_{j-1}}, 
\beta^{-1} \int_0^{\beta} ds
\mathbb{T}_{+g}
\big(
\omega_j(s)(\bfphi_j)
\big)e_{X_{j}}
\Bigg\ra
\Bigg|\no
\ge &
\gamma(\bphi, \bphi_1) \gamma(\bphi_1, \bphi_2)\cdots
 \gamma(\bphi_{L-1}, \bphi'),
\end{align} 
where $\gamma(\bphi, \bphi')$ is given by Lemma \ref{Tnonvanish}.
By combining this and (\ref{GTau}), we obtain the desired result. $\Box$

\end{document}